\documentclass[journal]{IEEEtran}
\usepackage{amssymb}
\usepackage{amsmath}
\usepackage{graphicx}
\newcommand{\B}{\mathfrak{B}}

\newcommand{\Hi}{\mathcal{H}}

\newcommand{\Tr}{\mathrm{Tr}}

\newcommand{\ket}[1]{| #1 \rangle}
\newcommand{\bra}[1]{\langle #1 |}
\newcommand{\braket}[2]{\langle #1 \vert #2 \rangle}
\newcommand{\be}{\begin{equation}}
\newcommand{\ee}{\end{equation}}
\newtheorem{Theorem}{Theorem}

\newtheorem{Lemma}{Lemma}
\newtheorem{Corollary}{Corollary}
\newtheorem{Proof}{Proof}

\ifCLASSINFOpdf
\else
\fi
\begin{document}
%
\title{Local hypothesis testing between a pure bipartite state and the
white noise state}
%
%
%

\author{Masaki~Owari,
        and~Masahito~Hayashi,
\thanks{M. Owari is with Institut f\"ur Theoretische Physik,
        Universit\"at Ulm.}
\thanks{M. Hayashi is with Division of Mathematics, Graduate School of Information Sciences 
Tohoku University, and
Centre for Quantum Technologies
National University of Singapore.}
}


\maketitle

\begin{abstract}
In this paper, we treat a local discrimination problem in the framework of
 asymmetric hypothesis testing. We choose a known bipartite pure state
 $\ket{\Psi}$ as an alternative hypothesis, and the completely mixed
 state as a null hypothesis.   As a result, we analytically derive an
 optimal type 2 error and  an optimal POVM for one-way LOCC POVM and
 Separable POVM. For two-way LOCC POVM, we study a family of
simple three-step LOCC protocols, and show that the best 
protocol in this family has strictly better performance than any one-way
LOCC protocol in all the cases where there may exist
difference between  two-way LOCC POVM and one-way LOCC POVM.
\end{abstract}

\IEEEpeerreviewmaketitle

\section{Introduction}
In all quantum information processings, we always need to measure
quantum states 
in order to derive classical information encoded there. 
Because of this, since an early stage of the field of quantum information, 
people have made effort to understand how well a given
unknown quantum state can be identified when a set of candidates
is given \cite{H76,H82}.
People deal this problem with different theoretical frameworks in 
the sub-fields of quantum information named, Quantum State
discrimination \cite{YKL75,PW91}, Quantum hypothesis testing
\cite{H72,HP91,ACMBMAV07}, Quantum State Estimation \cite{H67,YL73,H79}, and
Classical Capacity of Quantum Channel \cite{H79b,H98,SW97}
\footnote{Other references about these topic can be found  in the
reference lists of  \cite{H05,H06}.}.  

Because of decoherence,
we generally need to pay a lot of cost to reliably send a quantum state
to a spatially separated place.  
Thus, it is important to study quantum information processing
in a situation where 
reliable quantum communication is not available 
across spatially separated places;
this restriction for available quantum operations leads 
a class of quantum operations called 
LOCC (Local Operations and Classical Communication), 
and also other slightly different classes of quantum operations like
Separable Operations, PPT (Positive Partial Transpose) operations, 
etc \cite{BBPSSW96,BDSW96,VP07,HHHH09}.
Thus, many researches have been done 
to study how well a given partially unknown state can be identified
under these restricted quantum operations
\cite{BDFMRSSW99,WSHV00,VSPM01,GKRSS01,TDL01,DLT02,EW02,WH02,CY02,BHSS03,HM03,VP03,C04,F04,JCY05,HK05,N05,W05,HMMOV06,HMT06,OH06,O06,DFJY07,YJCZZG07,F07,MW08,OH08,XD08,HHH08,IHHH08,S08,SHHH09,DFXY09,BW09,H09,JRZZG09,N00}. 
These researches are often called
researches of ``{\it Local discrimination}''.
In this paper, we treat local discrimination in the framework of an asymmetric
hypothesis testing where we do not use any prior probability 
on a set of candidates.

In a hypothesis testing, we aim to certify a given hypothesis $H_1$
(called ``{\it alternative hypothesis}''), and
in order to do it, we try to reject a hypothesis $H_0 $ 
(called ``{\it null hypothesis}'')
which is true when $H_1$ is false. 
Hence, we try to minimize the error probability judging $H_0$
to be true when $H_1$ is true (the type 2 error)
under the condition that a fixed value $\alpha$ upper-bounds
the error probability judging $H_1$ to
be true when $H_0$ is true (the type 1 error).
When both $H_0$ and $H_1$ consist of a single state,
a hypothesis testing looks very similar to a normal state discrimination. 
However, they are different in the way to treat errors:
two kind of errors are treated in a completely asymmetric 
way in a hypothesis testing, 
and in a symmetric way in a state discrimination (although
their prior may not be symmetric).

The number of researches of an asymmetric quantum hypothesis testing 
is rather small with respect to that of quantum state discrimination;
a partial list of researches of asymmetric quantum hypothesis testing
may include \cite{H72,HP91,ON00,H02,OH04,HSTMTJ06,HMO07,NH07,HMO08,HMH09,M09}. 
In particular, concerning hypothesis testing with local restrictions
(we will called ``{\it local hypothesis testing}'' in this paper), 
only very restricted number of papers treated it \cite{HMT06,OH08,H09}. 

In this paper, we consider the situation where two spatially separated
parties detect a signal in a known bipartite pure state $\ket{\Psi}$.
They try to certify that what they detected is not a noise, but a state
$\ket{\Psi}$. On this purpose, we choose $\ket{\Psi}$ as an
alternative hypothesis and the completely mixed state, which represents
a white noise, as a null hypothesis. 
As a class of local measurements, we treat
one-way LOCC POVM (Positive Operator Valued Measure),
two-way LOCC POVM, and Separable POVM \cite{H06,VP07,HHHH09,NC00,DHR02}.  
This study can be considered as a
generalization of our previous paper \cite{OH08}; see Section \ref{sec
preliminary} for detail discussion about their relation. 

As a result, we analytically derive an optimal type 2 error and 
an optimal POVM for one-way LOCC POVM and Separable POVM. In particular, in
order to derive an analytical solution for Separable POVM, we   
prove the equivalence of the local hypothesis testing under
Separable POVM and a global hypothesis testing with a composite
null hypothesis, and analytically solve this global hypothesis 
testing.
Furthermore, for two-way LOCC POVM, we study a family of
simple three-step LOCC protocols, and show that the best 
protocol in this family has strictly better performance than any one-way
LOCC protocol in all the cases where there may exist
difference between  two-way LOCC POVM and one-way LOCC POVM.

In quantum information, so far, just a very limited number of works treat
a hypothesis testing with a composite hypothesis \cite{H09,N00,B10}.
In this paper, on the way to derive analytical solutions to the local
hypothesis testing under separable POVMs,
we add one example into this category. Our
example consists of a composite null hypothesis and a simple
alternative hypothesis on a single partite Hilbert-space.
A set of the null hypothesis is generated from a single pure state
by phase flipping operations. We give an analytical solution for this
global hypothesis testing with a composite null hypothesis.

This paper is organized as follows:
We explain notations and problem settings in Section \ref{sec
preliminary}, and, then, present main results of the paper in Section
\ref{sec main results}. One-way and two-way LOCC are treated in Section
\ref{sec locc}. We give an analytical solution for a global hypothesis
testing with a composite hypothesis in Section \ref{sec global}, and
then, prove the equivalence between this hypothesis testing and the
local hypothesis testing under Separable POVM in Section \ref{sec
separable}.  
Finally, we give a summary in Section \ref{sec summary}. We
also add appendix to present a proof for a corollary.

\section{Preliminary}\label{sec preliminary}
\subsection{Notations}
First, we introduce our notations. A finite bipartite Hilbert space is called as
$\Hi_{AB} \stackrel{\rm def}{=} \Hi _A \otimes \Hi _B$. We define $d_A$,
$d_B$ and $d$ as 
$d_A \stackrel{\rm def}{=} \dim \Hi _A$,  $d_B \stackrel{\rm def}{=} \dim
\Hi _B$ and $d \stackrel{\rm def}{=} \min \{ d_A, d_B \}$, respectively. 
Normally, we assume that two spatially separated parties, say Alice
and Bob, possess these two local Hilbert spaces $\Hi _A$ and $\Hi _B$, respectively.
The space of all operators on $\Hi$ is called $\B (\Hi)$.
The space of all Hermitian operators on $\Hi$ is called $ \mathcal{P}(\Hi)$.
The cone of all positive operators on $\Hi$ is called
$\mathcal{P}_+(\Hi)$. 
$\{ a < \rho \le b \}$ denotes a projection onto a direct sum of
eigenspaces whose eigenvalues $\lambda$ satisfy $a < \lambda < b$.

In this paper, we only consider a two-valued POVM $\{ T, I_{AB}-T \}$; 
$T \in \B (\Hi )$ satisfies $0 \le T \le I_{AB}$. 
Since a two-valued POVM is completely determined by fixing an element $T$,
we often say ``POVM $T$'' as an abbreviation of 
``POVM $\{ T, I-T\}$''  \footnote{We often abbreviate $I_{AB}$ as $I$ in
the case when it is apparent on which space $I$ is defined.} in this paper. 
A word ``global POVM'' just means a POVM with no additional restriction, 
and we denote a set of all two-valued POVMs on $\Hi_{AB}$ as $g$.
A POVM is called a two-way LOCC POVM, if it can be implemented by
two-way LOCC (local operations with two-way classical communication)
\cite{H06,VP07,HHHH09,DHR02}. 
$\leftrightarrow$ denotes a set of all two-values two-way LOCC. 
Similarly, a POVM is called a one-way LOCC POVM, if it can be implemented by
one-way LOCC (local operations with
one-way classical communication) \cite{H06,VP07,HHHH09,DHR02}.
There are two-different sets of one-way LOCC corresponding to 
two-different directions of one-way classical communication (C.C.);
that is, one-way LOCC with C.C. from Alice to Bob and 
it with C.C. from Bob to Alice. 
These two types of one-way LOCC should be treated distinctly.
However, in our case, the final results (an optimal error or success probability) corresponding to
one set can easily be derived from another by just swapping the dimension of Alice and Bob. 
We just treat a set of one-way LOCC POVMs from Alice to Bob,
and we write this set as $\rightarrow$. 
A POVM is called a separable POVM, if it can be implemented by a separable
operations \cite{H06,VP07,HHHH09,NC00,DHR02}. A POVM is separable if and only if all the elements are
separable \cite{VP03}: in this case, a POVM $\{ T, I-T \}$ is separable if and only if both $T$ and $I-T$ can be written as
\begin{align}
 T &= \sum _i A_i \otimes B_i \nonumber \\
 I - T &= \sum _i A'_i \otimes B'_i
\end{align}
by using positive operators $\{ A_i \}_i$, $\{ B_i \}_i$, 
$\{ A'_i \}_i$, and $\{ B'_i \}_i$.

\subsection{Problem Settings}
In this paper, we consider a hypothesis testing between a given fixed pure-bipartite
state $\ket{\Psi}$ and the completely  mixed state (or a white noise) $\rho _{mix}$ under the different
restrictions on available POVMs: global POVM, separable POVM, one-way LOCC POVM, two-way LOCC POVM. 
Especially, we consider the situation where we intend to assert that 
an unknown state is the pure-bipartite state $\ket{\Psi}$. 
In order to do so, we choose the completely mixed state $\rho_{mix}$ as
a null hypothesis and the state $\ket{\Psi}$ as an alternative hypothesis. 
That is, we minimize the error probability judging an unknown state to
be $\rho _{mix}$ when the state is actually $\ket{\Psi}$ (the type 2 error)
under the condition that a fixed value $\alpha$ upper-bounds
the error probability judging an unknown state to
be $\ket{\Psi}$ when the state is actually $\rho_{mix}$ (the type 1 error).

Our POVM consists of two POVM elements $T$ and $I-T$.
When the measurement result is $T$, we judge an unknown state as
$\ket{\Psi}$,
and when the measurement result is $I-T$, we judge an unknown state
as $\rho_{mix}$. 
Thus, the type 1 error is written as 
\begin{equation}
\alpha (T)= \Tr \rho_{mix}T,
\end{equation}
and the type 2 error is written as 
\begin{equation}
\beta (T) = \bra{\Psi}\left ( I-T \right ) \ket{\Psi}.
\end{equation}
As a result, the optimal type 2 error under the condition 
that the type 1 error is less than or equal to $\alpha$ is written as
\begin{eqnarray*}
\beta_{\ket{\Psi}, C}(\alpha)\stackrel{\rm def}{=}  \min _{T} \left \{
 \beta (T) \  | \ \alpha (T) \le \alpha, \{ T, I-T \} \in C \right \},
\end{eqnarray*}
where $C$ is either $\rightarrow$, $\leftrightarrow$, $Sep$, or $g$
corresponding to one-way LOCC, two-way LOCC, separable
POVM and the global POVM, respectively.
The optimal success probability
$S_{\alpha, C}(\ket{\Psi})$ is defined as
\begin{equation}
S_{\alpha, C}(\ket{\Psi})\stackrel{\rm def}{=}1-\beta_{\ket{\Psi}, C}(\alpha). 
\end{equation}
In this paper, we mainly try to derive the optimal type 2 error 
$\beta_{\ket{\Psi}, C}(\alpha)$ by 
calculating the optimal success probability $S_{\alpha, C}(\ket{\Psi})$, since the latter is slightly
simpler than the former.

We can easily calculate the optimal success probability for the global
POVM, which apparently does not depend on choice of the pure state $\ket{\Psi}$.
The result is 
\begin{eqnarray}
 \beta_{\ket{\Psi}, C}(\alpha)=d_Ad_B \min\{\alpha, 1/d_Ad_B\}.
\end{eqnarray} 
The optimal POVM is given by $T=\beta_{\ket{\Psi}, g}(\alpha)\ket{\Psi}\bra{\Psi}$.
Therefore, the purpose of this paper is evaluating 
$\beta_{\ket{\Psi}, C}(\alpha)$ for $C= \rightarrow, \leftrightarrow, Sep$,
and observing the trade-off 
between Type 1 error $\alpha$  and Type 2 error $\beta$.

\subsection{Swapping null and alternative hypotheses}
In this paper, we will mainly concern $\beta_{\ket{\Psi}, C}(\alpha)$ in this
paper. However, someone may be interested in the hypothesis testing 
whose null hypothesis and alternative hypothesis are the converses of
ours. The optimal type 2 error for this converse hypothesis testing 
corresponds to the the optimal type 1 error 
$\alpha_{\ket{\Psi},C}(\beta)$ for our problem under the
condition that type 2 error is less than a fixed value $\beta$:
\begin{eqnarray*}
 \alpha_{\ket{\Psi},C} (\beta) \stackrel{\rm def}{=}  \min _{T} \left \{
 \alpha (T) \  | \ \beta (T) \le \beta, \{ T, I-T \} \in C \right \}.
\end{eqnarray*} 
Since the trade-off $\beta_{\ket{\Psi},C}(\alpha)$ is a non-decreasing
function, the trade-off for the converse hypothesis testing 
$\alpha _{\ket{\Psi},C}(\beta)$ is given as
\begin{equation} \label{eq sec preliminary alpha c = min alpha}
 \alpha_{\ket{\Psi},C}(\beta)=\min \{ \alpha \ | \ \beta_{C, \ket{\Psi}}(\alpha)=\beta \}.
\end{equation} 
Especially, in the region where $\beta_{\ket{\Psi},C}(\alpha)$ is
strictly decreasing, it is given just as the inverse function of $\beta_{\ket{\Psi},C}(\alpha)$:
\begin{align}
 \alpha_{\ket{\Psi},C}(\beta)=\beta_{\ket{\Psi},C}^{-1}(\beta).
\end{align}
Actually, as we will see later, $\beta_{\ket{\Psi},C}(\alpha)$ is
strictly decreasing all the region of $\alpha$ except 
the region where $\alpha$ satisfies $\beta_{\ket{\Psi},C}(\alpha) =0$.
Therefore, the graph for the trade-off $\alpha_{\ket{\Psi},C}(\beta)$ is essentially
derived just by swapping the axes of the graph for the trade-off $\beta_{\ket{\Psi},C}(\alpha)$.

In the paper \cite{OH08}, we treated this converse hypothesis testing 
and derived the optimal type 2 error under the condition that the type 1
error is $0$. In our notation, it corresponds to 
$\alpha _{\ket{\Psi},C}(0)$.
Thus, the main results of \cite{OH08} can be written down as
\begin{align}
\alpha_{\ket{\Psi},Sep}(0)& = \frac{1}{d_Ad_B}(\Tr \sqrt{\rho_A})^2,\\
\alpha_{\ket{\Psi},\rightarrow}(0)& = \frac{1}{d_Ad_B}{\rm rank}\rho_A,
\end{align}
and 
\begin{align}
\quad& \alpha_{\ket{\Psi},\leftrightarrow}(0)\nonumber \\
\le & 
\frac{1}{d_Ad_B} \min_{\{\delta_{ki}\}_{1 \le k \le i \le d}}
\Big \{ \sum_{i=1}^{d}i \cdot \frac{\sum_{k=1}^i \lambda _k
\delta_{ki}^2}{
\sum_{k=1}^i \lambda _k \delta_{ki}}
\ \Big | \nonumber \\
\quad& \qquad \forall k, \forall i, \delta_{ki} \ge 0 \ {\rm and}\  \forall k,
\sum_{i=k}^{d} \delta_{ki}=1
\Big \},
\end{align}
where $d$ is defined as $d \stackrel{\rm def}{=}\min \{d_A, d_B\}$, and 
$\{ \lambda _k \}_{k=1}^{d}$ is the Schmidt coefficients of $\ket{\Psi}$
satisfying $\lambda_k \le \lambda_{k+1}$ for all $k$.
Therefore, from Eq.(\ref{eq sec preliminary alpha c = min alpha}), 
we have already known the smallest zero of $\beta_{\ket{\Psi},C}(\alpha)$.

\section{Main results}\label{sec main results}
In this section, we give the main results of this paper. 
In the following parts, we always choose computational basis 
as the Schmidt basis of $\ket{\Psi}$ in the following way:
\begin{equation}
 \ket{\Psi}=\sum _{i=1}^{d}\sqrt{\lambda_i}\ket{ii},
\end{equation}
where $d \stackrel{\rm def}{=} \min \{ d_A, d_B \}$ and 
$\{ \lambda \}_{i=1}^d$ are the Schmidt coefficients of $\ket{\Psi}$
satisfying $\lambda_{i} \ge \lambda _{i+1}$.

For one-way LOCC POVM, 
we prove that an optimal strategy is measuring an unknown state in each local
computational basis and post-processing the measurement results.
Thus,
the local hypothesis testing under one-way LOCC is essentially equivalent to a
classical hypothesis testing between a probability distribution defined
by the Schmidt coefficients of $\ket{\Psi}$ and the classical white noise
(Lemma \ref{lemma sec one-way additional}).
As a result, the optimal type 2 error is given by the
following theorem:
\begin{Theorem}\label{main theorem one-way}
 Defining a natural number $c$ as 
\begin{equation}\label{eq sec one-way locc def c}
 c \stackrel{\rm def}{=} \min \left \{d,  \left \lfloor d_Ad_B\alpha\right \rfloor +1
						     \right \},
\end{equation}
then, for a state $\ket{\Psi}= \sum _i \sqrt{\lambda_i}\ket{ii}$ with
 $\lambda_i \ge \lambda_{i+1}$, 
$\beta_{\ket{\Psi}, \rightarrow}(\alpha)$ can be written as
\begin{eqnarray}
 \beta_{\ket{\Psi}, \rightarrow}(\alpha) = \sum _{i=c}^{d}\lambda_i
- m_c\lambda_c,
\end{eqnarray}
where $m_c$ is defined as 
\begin{equation}\label{eq sec one-way locc def m c}
 m_c \stackrel{\rm def}{=} \min \{1, d_Ad_B\alpha -c+1 \}.
\end{equation}
An optimal POVM can be written as $\{T, I-T \}$ by using the following
 $T \in \B (\Hi_{AB})$: 
\begin{equation}\label{eq sec main def optimal one-way locc povm}
 T=\sum_{i=1}^{c-1}\ket{ii}\bra{ii} + m_c\ket{cc}\bra{cc}.
\end{equation}
\end{Theorem}

Since the definition of two-way LOCC is mathematically complicated in comparison to that of
one-way LOCC and separable operations \cite{H06,OH08,DHR02},  
it is extremely difficult to evaluate the optimal error probability
for two-way LOCC POVM.
Therefore, we only evaluate performance of a particular type of two-way LOCC protocols
which belong to three steps LOCC and are used in the previous paper
\cite{OH08}.
Hence, we only derive an upper bound for the optimal type 2
error for 2-way LOCC: Defining $\widetilde{\beta}_{\ket{\Psi}, \leftrightarrow}(\alpha)$ as
\begin{align}\label{eq sec main result def overline beta}
\quad& \widetilde{\beta}_{\ket{\Psi}, \leftrightarrow}(\alpha) \nonumber \\
\stackrel{\rm def}{=}&1-\max _{ \{ m_{i}^k \}_{1\le k \le i \le d_A}}\Big \{ \sum _{i, k} \lambda_k
 m_i^k  \  \Big | \  0 \le m_i^k,  \nonumber\\ 
\quad& \sum_{i=k}^{d_A} m_i^k \le 1, \sum _{i=1}^{d_A} i\cdot
\frac{\sum _{k = 1}^{i}\lambda_k (m_i^k)^2}{\sum _{k = 1}^{i}\lambda_km_i^k }\le \alpha
d_Ad_B \Big  \}, \nonumber \\
& \quad  
\end{align}
we derive the following theorem:
\begin{Theorem}\label{main theorem two-way}
\begin{equation}
  \widetilde{\beta}_{\ket{\Psi}, \leftrightarrow}(\alpha) \ge \beta_{\ket{\Psi},
 \leftrightarrow}(\alpha).
\end{equation}
\end{Theorem}
This upper bound $\widetilde{\beta}_{\ket{\Psi}, \leftrightarrow}(\alpha)$
is in the form of a convex optimization with 
$\frac{d_A\left ( d_A+ 1 \right )}{2}$ parameters.

For separable POVMs, we prove that this local hypothesis testing problem
is equivalent to another hypothesis testing problem
with a composite null hypothesis under global POVM, and by solving this
simpler hypothesis testing problem, we derive an optimal type 2 error
for the original local hypothesis testing problem. 
Here, we only give the final theorem for the local hypothesis testing
under separable POVM.
First, we can assume $d_A \le d_B$ without losing generality.
For given $\alpha >0$ and $\ket{\Psi}$, we introduce the following notations:
For a natural number $l \le d_A$, a real number $\epsilon_l$ is defined as $\epsilon _l \stackrel{\rm def}{=}
 \sqrt{\frac{\alpha d_Ad_B}{l}}$,
a state $\ket{\psi_l}$ is defined as \begin{equation}\label{eq sec main results def ket psi_l}
 \ket{\psi_l} \stackrel{\rm def}{=}
\sum _{i=1}^l \sqrt{\lambda_i}\ket{i}/\sqrt{\sum_{i=1}^l \lambda_i},
\end{equation} 
a state
 $\ket{\phi_l}$ is defined as
\begin{equation} \label{eq sec main results def phi_d}
\ket{\phi _{l}}=  \frac{1}{\sqrt{l}}\sum _{i=1}^l \ket{i},
\end{equation}
 and a
 state $\ket{\phi'_l}$ is defined as
\begin{equation} \label{eq sec main results theorem 1 optimal state non-trivial}
 \ket{\phi'_l}=\frac{\sqrt{1-\epsilon _{l}^2}\ket{\psi_{l}}- \left (
c_{l}\sqrt{1-\epsilon_{l}^2}-\epsilon_{l} \sqrt{1-c_{l}^2} \right )\ket{\phi_{l}}}{\sqrt{1-c_{l}^2}},
\end{equation}
where $c_l$ is defined as $c_l\stackrel{\rm def}{=}
\braket{\psi_l}{\phi_l}$.
By using the above notations, a natural number $\eta$ is defined as 
\begin{align}
\eta \stackrel{\rm def}{=} 
\left \{
\begin{array}{l}
d_A  \quad {\rm if} \ \epsilon _{d_A} \ge
 \braket{\phi_{d_A}}{\psi_{d_A}} \ {\rm or} \
\ket{\psi_{d_A}} =  \ket{\phi _{d_A}}
 \\
\quad \\
{\rm otherwise} \\
\min _{l \in \mathbb{N}} \Big \{ l \  \Big
 | \ l \le d,
 \epsilon _l < \braket{\phi_l}{\psi_l},   \\
\quad \qquad \qquad  \ket{\psi _l} \neq  \ket{\phi
 _l}, \braket{l}{\phi'_l} <0 \Big \} -1
\end{array}
\right .
\end{align}
By the definition, $\eta$ satisfies $1 \le \eta \le d_A$.
Further, we define an operator $T(\ket{\phi})$ depending on a vector 
$\ket{\phi} \in \Hi_A$ as 
\begin{eqnarray*}
 T(\ket{\phi})\stackrel{\rm def}{=}V\ket{\phi}\bra{\phi}V^{\dagger}
+\sum_{j\neq
k}\sqrt{\braket{j}{\phi}\braket{\phi}{k}}\ket{j}\bra{j}\otimes 
\ket{k}\bra{k}.
\end{eqnarray*}
In the above formula, $V$ is an isometry
 between $\Hi_A$ and $\Hi_{AB}$  defined as 
$V \stackrel{\rm def}{=}\sum_i \ket{ii}\bra{i}$.
As we will prove later, $\{ T(\ket{\phi}), I-T(\ket{\phi})\}$ is
a separable POVM for all $\ket{\phi} \in \Hi _A$. 
Then, by using the above notations, the optimal type 2 error is given by the following
theorem:
\begin{Theorem} \label{main theorem sep}
\begin{enumerate}
 \item In the case when $\epsilon _{\eta} \ge  \braket{\phi_{\eta}}{\psi_{\eta}}$, 
\begin{equation}\label{eq sec sep theorem 1 solution trivial}
\beta_{\ket{\Psi}, Sep}(\alpha)=1 - \sum _{i=1}^{\eta} \lambda_i,
\end{equation}
and a POVM $T(\ket{\psi_{\eta}})$ attains the optimum.

\item In the case when $\epsilon_{\eta} <    \braket{\phi_{\eta}}{\psi_{\eta}}$,
\begin{equation}\label{eq sec sep theorem 1 solution non-trivial}
 \beta_{\ket{\Psi}, Sep}(\alpha)=  1 - \left (\sum_{i=1}^{\eta}\lambda_i \right)\cdot 
\left ( \sqrt{1-\epsilon_{\eta}^2} \sqrt{1-c_{\eta}^2} +
 \epsilon_{\eta} c_{\eta} \right )^2.
\end{equation}
A POVM 
$T(\ket{\phi'_{\eta}})$ 
attains the optimum in the case 
$\ket{\psi_{\eta}} \neq \ket{\phi  _{\eta}}$,
a POVM 
$T(\epsilon_1 \ket{1})$ 
attains the optimum in the case $\eta=1$,
and a POVM 
$T(\epsilon_{\eta}\ket{\phi_{\eta}}+
      \sqrt{1-\epsilon_{\eta}^2}\ket{\phi_{\eta}^{\perp}} )$ 
attains the optimum in the case $\eta \ge 2$ and 
$\ket{\psi_{\eta}} = \ket{\phi  _{\eta}}$.
Here, $\ket{\phi_{\eta}^{\perp}}$ is any state orthogonal to
$\ket{\phi_{\eta}}$ and, thus, can be chosen as 
$\ket{\phi_{\eta}^{\perp}}=(\ket{1}-\ket{2})/\sqrt{2}$.
\end{enumerate}
\end{Theorem}

Before discussing plots of $\beta_{\ket{\Psi}, C}(\alpha)$, 
we explain several facts which can be easily seen from the above main
theorems.
For the global POVM, we can trivially derive $\beta_{\ket{\Psi}, g}(\alpha)=0$ for $\alpha \ge 1/d_Ad_B$.
On the other hand, for the other local POVMs, we derive 
$\beta_{\ket{\Psi}, sep}(\alpha)=\beta_{\ket{\Psi},\leftrightarrow}(\alpha)= \beta_{\ket{\Psi}, \rightarrow}(\alpha)=0$ 
for $\alpha \le 1/\max \{d_A,d_B\}$.
The latter can be easily seen from Theorem \ref{main theorem one-way}.
Moreover, we can derive the following corollary from the above theorem:
\begin{Corollary}\label{corollary sec main}
 For $\alpha < 1/d_Ad_B$, 
\begin{equation}\label{eq sec main beta = beta =beta}
 \beta_{\ket{\Psi},
  sep}(\alpha)=\beta_{\ket{\Psi},\leftrightarrow}(\alpha)=
  \beta_{\ket{\Psi}, \rightarrow}(\alpha) =1-\lambda_1\alpha d_Ad_B.
\end{equation}
The optimal POVM is given by $T=1-\lambda_1 \alpha d_Ad_B$. 
When $\ket{\Psi}$ is a product state or a maximally entangled state, 
Eq.(\ref{eq sec main beta = beta =beta}) holds all the region $0 \le
 \alpha \le 1/\max \{d_A,d_B\}$
\end{Corollary}
\begin{Proof}
See Appendix A.
\end{Proof}
Thus, separable and one-way and two-way LOCC POVM just give the same
optimal error for $\alpha <1/d_Ad_B$ and $\alpha > 1/\max \{d_A, d_B \}$
for a non-maximally entangled state $\ket{\Psi}$. On the other hand,
they just coincide in all the region for a maximally entangled state $\ket{\Psi}$.

\begin{figure}[!t]
\centering
\includegraphics[width=3.0in]{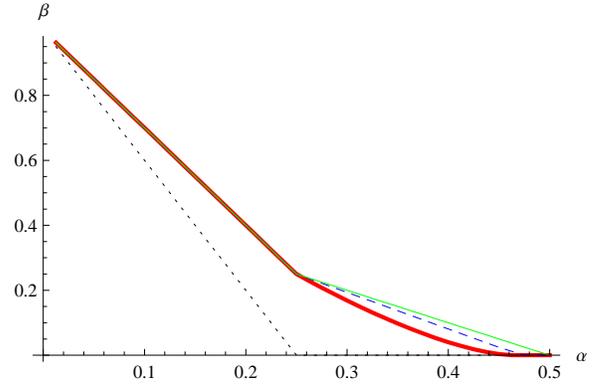}
\caption{The trade-off between the type 1 error ($\alpha$) and the type 2 error
 ($\beta$) for 
$\ket{\Psi}=\frac{\sqrt{3}}{2}\ket{11}+\frac{1}{2}\ket{22}$. 
Thin line: $\beta_{\ket{\Psi},\rightarrow}(\alpha)$; Broken line:
 $\widetilde{\beta}_{\ket{\Psi},\leftrightarrow}(\alpha)$; Thick line:
 $\beta_{\ket{\Psi},sep}(\alpha)$; Dotted line:  
$\beta_{\ket{\Psi},g}(\alpha)$.}
\label{fig rho13}
\end{figure}
\begin{figure}[!t]
\centering
\includegraphics[width=3.0in]{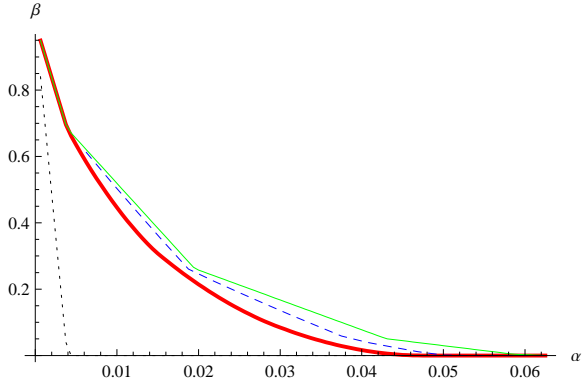}
\caption{The trade-off between the type 1 error ($\alpha$) and the type 2 error
 ($\beta$) for 
$\ket{\Psi}=\{\frac{\sqrt{3}}{2}\ket{11}+\frac{1}{2}\ket{22}\}^{\otimes
 4}$
Thin line: $\beta_{\ket{\Psi},\rightarrow}(\alpha)$; Broken line:
 $\widetilde{\beta}_{\ket{\Psi},\leftrightarrow}(\alpha)$; Thick line:
 $\beta_{\ket{\Psi},sep}(\alpha)$; Dotted line:  
$\beta_{\ket{\Psi},g}(\alpha)$.}
\label{fig rho13quad}
\end{figure}
Now, we present several figures about graphs of the trade-off between the type 1 error
$\alpha$ and the type 2 error $\beta$ for global, separable, two-way
LOCC, and one-way LOCC POVM. 
For two-way LOCC POVM, we draw the graph of
$\tilde{\beta}_{\ket{\Psi}, \leftrightarrow}(\alpha)$ 
instead of $\beta_{\ket{\Psi},\leftrightarrow}(\alpha)$.
Here, we always choose
$d_A=d_B=d$ for simplicity. First, we give graphs of the trade-off for
$\ket{\Psi}=\frac{\sqrt{3}}{2}\ket{11}+\frac{1}{2}\ket{22}$
(FIG. \ref{fig rho13}) and 
$\ket{\Psi}=\left (\frac{\sqrt{3}}{2}\ket{11}+\frac{1}{2}\ket{22}\right
)^{\otimes 4}$ (FIG. \ref{fig rho13quad}).
The graphs for separable, one-way LOCC and two-way
LOCC coincide in the regions  $\alpha \le 1/d^2$ and $\alpha \le 1/d$.
On the other hands, they separate in all the region 
$1/d^2 < \alpha < 1/d$, that is, 
$\beta_{\ket{\Psi},sep}(\alpha)$ is strictly smaller than
$\tilde{\beta}_{\ket{\Psi},\leftrightarrow}(\alpha)$,
and also $\tilde{\beta}_{\ket{\Psi},\leftrightarrow}(\alpha)$ 
is smaller than $\beta_{\ket{\Psi},\rightarrow}(\alpha)$.
  In the previous paper \cite{OH08}, we observed
improvement of the optimal error probability from one-way (two-steps) LOCC to
three-steps two-way LOCC by using
the same simple three-steps LOCC protocol used in this paper 
for $\tilde{\beta}_{\ket{\Psi},\leftrightarrow}(\alpha)$.
As we have explained, these optimal error probabilities in the previous
paper correspond to the smallest zeros of the graphs
$\tilde{\beta}_{\ket{\Psi},\leftrightarrow} (\alpha)$ and
$\beta_{\ket{\Psi},\rightarrow}(\alpha)$ in the present paper. 
The presented graphs show that the similar improvement is observed all
the region of $1/d^2 < \alpha < 1/d$.
\begin{figure}[!t]
\centering
\includegraphics[width=3.0in]{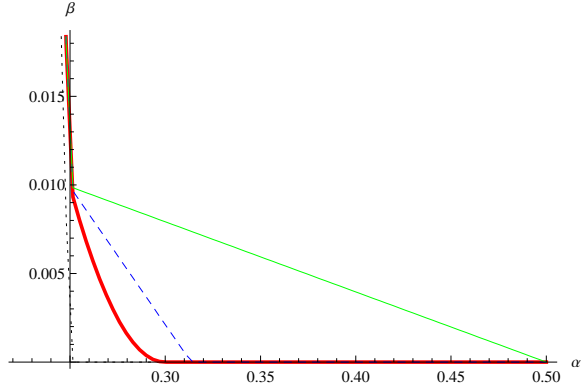}
\caption{The trade-off between the type 1 error ($\alpha$) and the type 2 error
 ($\beta$) for 
$\ket{\Psi}=\frac{10}{\sqrt{101}}\ket{11}+\frac{1}{\sqrt{101}}\ket{22}$
Thin line: $\beta_{\ket{\Psi},\rightarrow}(\alpha)$; Broken line:
 $\widetilde{\beta}_{\ket{\Psi},\leftrightarrow}(\alpha)$; Thick line:
 $\beta_{\ket{\Psi},sep}(\alpha)$; Dotted line:  
$\beta_{\ket{\Psi},g}(\alpha)$.}
\label{fig rho1100}
\end{figure}
As we can observe from FIG \ref{fig rho1100}, 
when $\ket{\Psi}$ just have small entanglement, this improvement can be
seen more clearly.
In other words, in this case, 
the straight line $\tilde{\beta}_{\ket{\Psi},\leftrightarrow}(\alpha)$ 
gives an approximation of the curve $\beta_{\ket{\Psi},sep}(\alpha)$ in
the region $1/d^2 < \alpha < 1/d$. 
\begin{figure}[!t]
\centering
\includegraphics[width=3.0in]{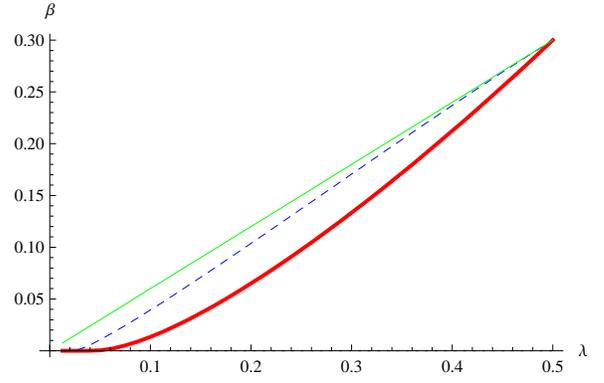}
\caption{Type 2 error as a function of $\lambda$ with $\alpha=0.35$, 
where $\lambda$ is
 defined as $\ket{\Psi}= \sqrt{\lambda}\ket{00}+ \sqrt{1-\lambda}\ket{11}$.
Thin line: $\beta_{\ket{\Psi},\rightarrow}(\alpha)$; Broken line:
 $\widetilde{\beta}_{\ket{\Psi},\leftrightarrow}(\alpha)$; Thick line:
 $\beta_{\ket{\Psi},sep}(\alpha)$.}
\label{fig delta35}
\end{figure}
Finally, we give a graph showing the variation with $\ket{\Psi}$ of 
$\beta_{\ket{\Psi},C}(\alpha)$ for a fixed $\alpha$ 
(FIG \ref{fig delta35}). As we have explained in Corollary
\ref{corollary sec main}, the graphs coincide when $\ket{\Psi}$ is a
product state ($\lambda=0$) and a maximally entangled state
($\lambda=0.5$).
On the other hand, the difference between
$\beta_{\ket{\Psi},\rightarrow}(\alpha)$ and 
$\tilde{\beta}_{\ket{\Psi},\leftrightarrow}(\alpha)$ is maximized when
$\beta$ is closed to $0$.

\section{Hypothesis testing under LOCC}\label{sec locc}
In this section, we treat the hypothesis testing under LOCC.
In the first subsection, we treat one-way LOCC 
and give a proof of Theorem \ref{main theorem one-way}. 
In the second subsection, we give a detail discussion about two-way LOCC
protocols including a proof of Theorem \ref{main theorem two-way}.

\subsection{One-way LOCC}
The main purpose of this subsection is giving a proof of 
Theorem \ref{main theorem one-way}, which gives an optimal type 2 error 
probability $\beta_{\ket{\Psi}, \rightarrow}(\alpha)$ under one-way LOCC.

When we consider one-way LOCC  on a bipartite system \cite{OH08,DHR02,H06}, 
there are two possibilities for a direction of classical communications, 
that is, from Alice $\Hi_A$ to Bob $\Hi_B$ and from Bob $\Hi_B$ to Alice
$\Hi _A$.  Here, since a state $\ket{\Psi}$ is symmetric under the
swapping between Alice and Bob,  we can
restrict ourselves into the situation where Alice send a
classical message to Bob without losing generality. 
Thus, we are interested in an optimal
success probability $S_{\alpha, \rightarrow}$ defined as
 \begin{align}
 \quad&  S_{\alpha, \rightarrow}(\ket{\Psi}) \nonumber \\
 \stackrel{\rm def}{=}&  \max _{T} \left \{
 \bra{\Psi}T\ket{\Psi} | \Tr \rho_{mix}T \le \alpha, \{ T, I-T \} \in \rightarrow \right \},
 \end{align}
 where $\rightarrow$ is a set of all one-way LOCC POVMs. 

 We first derive the following lemma, which reduces our local hypothesis testing problem to
a hypothesis testing problem defined just on a single Hilbert space:
\begin{Lemma}
 \begin{align}\label{eq sec one-way S delta rightarrow = max M }
\quad& S_{\alpha, \rightarrow}(\ket{\Psi}) \nonumber\\
=&  \max _{M \in \B (\Hi_A)} \Big \{
\Tr \rho _A M \ | \ \Tr M \le d_A d_B \alpha, 
 0 \le M \le I_A \Big \},
 \end{align}
where $\rho_A$ is a reduced density matrix of $\ket{\Psi}$; $\rho_A
 \stackrel{\rm def}{=} \Tr _B \ket{\Psi}\bra{\Psi}$.
\end{Lemma}
\begin{Proof}
 Without losing generality, we can choose Alice's POVM 
as a rank one POVM. Thus, an optimal POVM can be written as $T=\sum _m
 \ket{m}\bra{m}\otimes N_m$, where $\sum_m \ket{m}\bra{m}=I_A$ and $0
 \le N _m \le I_B$.
After Alice's measurement, Bob's system $\Hi_B$ is in a state
 $\ket{p_m}\stackrel{\rm def}{=} \braket{m}{\Psi}/\|\braket{m}{\Psi}\|$.
Suppose $T'$ is defined as 
\begin{equation}
 T'\stackrel{\rm def}{=} \sum _m \bra{p_m}N_m\ket{p_m}\ket{m}\bra{m}\otimes \ket{p_m}\bra{p_m}.
\end{equation}
\end{Proof}
Then, this new one-way LOCC POVM $T'$ satisfies
$\bra{\Psi}T\ket{\Psi}=\bra{\Psi}T'\ket{\Psi}$ and $\Tr T' \le \Tr T$.
Thus, $T'$ is also an optimal POVM.
Defining $M \stackrel{\rm def}{=} \sum _m \|\ket{m}\|^2 \cdot
\bra{p_m}N_m\ket{p_m}$,
we have $\Tr T'= \Tr M$. 
Moreover, $\bra{\Psi}T'\ket{\Psi}$ can be evaluated as
\begin{align}
 \bra{\Psi}T'\ket{\Psi} =& \sum _m \| \braket{m}{\Psi}\|^2 \cdot
  \bra{p_m}N_m\ket{p_m} \nonumber \\
=& \Tr \rho_A M.
\end{align}
$0 \le \bra{p_m}N_m\ket{p_m} \le 1$ and $\sum _m \ket{m}\bra{m}=I_A$
guarantees $0 \le M \le I_A$.
Therefore, we derive 
 \begin{align}\label{eq sec one-way S delta rightarrow le max M }
\quad& S_{\alpha, \rightarrow}(\ket{\Psi}) \nonumber \\
\le&  \max _{M \in \B (\Hi_A)} \Big \{
\Tr \rho _A M | \Tr M \le d_A d_B \alpha, 0 \le M \le I_A \Big \}.
\end{align}
On the other hand, suppose an operator $M$ attains the optimum of the
right-hand side of the above inequality, and has a spectral
decomposition $M = \sum _m q_m \ket{m}\bra{m}$.
By defining a one-way LOCC POVM element $T$ as 
$T \stackrel{\rm def}{=} \sum _m
q_m \ket{m}\bra{m} \otimes \ket{p_m}\bra{p_m}$, where 
$\ket{p_m} \stackrel{\rm def}{=}\braket{m}{\Psi}/\|\braket{m}{\Psi}\|$,
we can easily see that this POVM element attains 
Eq.(\ref{eq sec one-way S delta rightarrow le max M }).
Therefore, we derive Eq.(\ref{eq sec one-way S delta rightarrow = max M }).
\hfill $\square$

We further reduce $S_{\alpha, \rightarrow}(\ket{\Psi})$ as follows:
\begin{Lemma}\label{lemma sec one-way additional}
\begin{align}\label{eq sec one-way additional}
\quad& S_{\alpha, \rightarrow}(\ket{\Psi}) \nonumber \\
=& \max \Big \{ \sum _{i=1}^{r} \lambda_im_i
\ \Big  | \
\sum _{i=1}^{d_A}m_i \le d_Ad_B\alpha,  0 \le m_i \le 1 \Big \}.
\end{align} 
\end{Lemma}
\begin{Proof}
  By the definition, $\rho _A$ can be written as 
 $\rho _A = \sum _i \lambda_i\ket{i}\bra{i}$.
 Suppose $M$ is optimal. 
 Then, we can defined a new operator $M'$ by means of pinching as 
 $M'=\sum_i \bra{i}M\ket{i}\ket{i}\bra{i}$. 
 It is straightforward to check $\Tr M'=\Tr M$, 
 $\Tr \rho_A M'=\Tr \rho_A M$, and $0 \le M' \le I_A$.
 Thus, $M'$ is also optimal.
 Hence, we can always choose an optimal $M$ as $M=\sum _i m_i
 \ket{i}\bra{i}$.
 Thus, we derive Eq.(\ref{eq sec one-way additional}). 
\hfill $\square$
\end{Proof}
This lemma show that the local hypothesis testing under one-way LOCC is
essentially equivalent to a hypothesis testing of two classical probability
distributions: $\{ \lambda_i \}_{i=1}^{d_A}$ and $\{ 1/d_A \}_{i=1}^{d_A}$.

By means of the above lemma, we can 
give a proof of Theorem \ref{main theorem one-way}, which
gives an analytical solution for the
hypothesis testing under one-way LOCC as follows:
\begin{Proof}[Theorem \ref{main theorem one-way}]
From the above lemma, we can choose $m_i=0$ for all $i > r$.
In the case when $r \le d_Ad_B\alpha$, 
an optimum is attained when $m_i=1$ for all $0 \le i \le r$,
and we have $S_{\alpha, \rightarrow}(\ket{\Psi})=1$.
Thus, we only consider the case when $r > d_Ad_B\alpha$ in the following
 part.

Suppose $\{m_i\}_{i=1}^{r}$ attains the optimum. 
First, we prove $\sum _{i=1}^rm_i=d_Ad_B\alpha$ by contradiction.
Suppose $\sum_{i=1}^rm_i < d_Ad_B\alpha$.
Then, there exists $i_0$ such that $m_{i_0} <1$.
Thus, there exists $\epsilon >0$ such that 
$m_{i_0}+\epsilon \le 1$.
By defining $m'_{i_0}=m_{i_0}+\epsilon$ and 
$m'_i=m_i$ for all $i \neq i_0$, $\{ m'_i \}_{i=1}^r$ satisfies
$\sum_{i=1}^r\lambda_im'_i > \sum _{i=1}^r\lambda_im_i$.
Thus, $\{ m_i\}_{i=1}^r$ is not optimal.
This is contradiction. 
Therefore, $\sum _{i=1}^rm_i=d_Ad_B\alpha$.

Second, we prove that an optimal $\{ m_i \}_{i=1}^r$
satisfies $m_i=1$ for all $i \le c-1$ and $m_i=0$ for all $i > c$ by
 contradiction.

For an optimal $\{ m_i \}_{i=1}^r$, suppose there exists a pair of natural numbers $k$ and $l$
 such that $k < l \le r$, $m_k<1$ and $m_l>0$.
Then, by defining $\{ m'_i \}_{i=1}^r$ as 
$m'_k \stackrel{\rm def}{=} \min \{ 1, m_k+m_l \}$, 
$m'_l \stackrel{\rm def}{=} \max \{ 0, m_l -(1-m_k) \}$,
and $m'_i=m_i$ for all $i$ satisfying $i \neq k$ and $i \neq l$,
we derive $\sum _{i=1}^rm'_i=d_Ad_B\alpha$. 
We have $\sum _{i=1}^r\lambda_im'_i > \sum _{i=1}^r \lambda_im_i$
 for $\lambda_k > \lambda_l$, and $\sum _{i=1}^r\lambda_im'_i = \sum _{i=1}^r \lambda_im_i$ for $\lambda_k=\lambda_l$.
Thus, when $\lambda_k > \lambda_l$, this is contradiction,
and when $\lambda_k=\lambda_l$, $\{ m'_i\}_{i=1}^r$ also gives an optimal POVM.  
Thus, when $k<l \le r$ and $\lambda_k > \lambda_l$, we have either $m_k=1$
 or $m_l=0$. 
Therefore, there exist $c_1$ and $c_2$ such that an optimal $\{ m _i \}_{i=1}^r$ satisfies $m_i=1$ for all
 $i < c_1$, $m_i=0$ for all $i \ge c_2$, and $\lambda_{c_1}= \cdots =\lambda_{c_2-1}$.
Thus, suppose $\{ m_i \}_{i=1}^r$ is optimal.
$\{ m'_i \}_{i=1}^r$ is also optimal when it
 satisfies $m_i=1$ for all
 $i < c_1$, $m_i=0$ for all $i \ge c_2$, and
 $\sum_{c_1}^{c_2-1}m'_i=\sum_{c_1}^{c_2-1}m_i$.
Especially, we can choose an optimal $\{ m_i \}_{i=1}^r$ as one satisfying
 $m_i=1$ for all $i < c$, $m_i=0$ for all $i \ge c+1$ for $c$ defined by
\begin{equation}
 c \stackrel{\rm def}{=} \max _{n \in \mathbb{Z}_+} \left \{n \ |\ n \le
						     d_Ad_B \alpha
						     \right  \} +1.
\end{equation}
In this case $m_c$ can be written down as 
\begin{equation}
 m_c \stackrel{\rm def}{=}d_Ad_B\alpha-c+1.
\end{equation}
\hfill $\square$
\end{Proof}

Finally, Theorem \ref{main theorem one-way}, the optimal 1-way
LOCC strategy can be described as follows:  
Alice and Bob independently measure their system in the Schmidt basis.
When they get the measurement result $\ket{ii}$ for $i \le c-1$, they judge 
the given state to be $\ket{\Psi}$. When  they get $\ket{cc}$, they conclude
$\ket{\Psi}$ in the probability $1-m_c$, and in all other cases, they
judge the state to be $\rho _{mix}$.

\subsection{Two-way LOCC}
In this subsection, we treat the hypothesis testing under the
restriction of two-way LOCC.  
The definition of two-way LOCC is mathematically complicated in comparison to that of
one-way LOCC and separable operations \cite{H06,OH08,DHR02}. 
Hence,  it is extremely difficult to evaluate 
optimal performance of information processing under the restriction of
two-way LOCC except in the case when we only concern the first exponent
of asymptotics (see Section 3.5 of \cite{H06}), or 
when we can prove the optimal performance with two-way LOCC is the same as that with one-way
LOCC, or separable operations.
Therefore, we only evaluate performance of a particular type of two-way LOCC protocols
belonging to three steps LOCC by a numerical optimization.

Suppose a bipartite state $\ket{\Psi} \in \Hi_{AB}$ is shared by
Alice $(\Hi_A)$ and Bob $(\Hi_B)$. Then, without losing generality, we can assume that a given three-steps protocol
consists of the first Alice's measurement $\{ M_i \}_{i \in I}$, the first Bob's
measurement $\{ N_j^i \}_{j \in J}$ depending on the first Alice's measurement results $i$, and
the second Alice's measurement $\{L^{ij},I_A-L^{ij}\}$ depending on the
first Alice and Bob's measurement results $i$ and $j$.
If the first Alice and Bob's measurement results satisfy $i \in I_0 \subset I$
and $j \in J_0 \subset J$, and Alice gets $L^{ij}$ as the second
measurement result, she judges that the given state is $\ket{\Psi}$,
and otherwise, she judges that it is $\rho_{mix}$.
Thus, we can write down a POVM element corresponding to $\ket{\Psi}$ as 
\begin{eqnarray}
 T= \sum _{i \in I_0, j \in J_0}\sqrt{M_i}L^{ij}\sqrt{M_i} \otimes N_{i}^j,
\end{eqnarray}
where $0 \le \sum _i M_i \le I_A$, $0 \le \sum_j N_i^j \le I_B$, and 
$0 \le L^{ij} \le I_A$. 
Without losing generality, we can assume that Bob never judges whether a
given state is $\ket{\Psi}$ or $\rho_{mix}$; that is, Alice makes all
decisions. 
Then, since Bob's state after Alice's first measurement can be written
down as $\{ (\sqrt{\rho_A}M_i\sqrt{\rho_A}) >0 \}$, an optimal Bob's
measurement can satisfy $\sum _{j \in J_0}N_j^i = \{ (\sqrt{\rho_A}M_i\sqrt{\rho_A})
>0 \}$,
where $\{ X > 0 \}$ is an orthogonal projection to the subspace spanned
by all eigen vectors of $X$ corresponding to strictly positive eigen
values \cite{H06}.

An optimal success probability $\bra{\Psi}T\ket{\Psi}$ under the above
restrictions is still too complicated to get the value by numeric. 
Even in the case $d_A=d_B=2$, the
optimization problem is non-convex nonlinear programming including
unlimited number of parameters.  Thus, here, we only consider a
particular type of protocols which are derived from the three step LOCC
protocol used in \cite{OH08} by small modifications.
The protocol is derived by means of the following two restriction from
general three step LOCC protocols. 
\begin{enumerate}
 \item As a first assumption, we choose
       $\ket{\xi_j^i}\bra{\xi_j^i}$ as Bob's measurement $N_j^i$, 
where $\{ \ket{\xi _j^i}\}_{j=1}^{{\rm rank}M_i}$ is a mutually unbiased
       basis for the eigen basis of Bob's state after Alice's first
       measurement \cite{WF89}.
It is known that Bob can send all the quantum information of his system to Alice 
by  measurements in a  mutually unbiased basis when Alice and Bob's
       system is in a pure state \cite{VSPM01,OH06}.
Since when a given state is $\ket{\Psi}$, the state after Alice's
       measurement is a pure state, Bob can send all the information for
       Alice in this case. 

\item Second, we assume that in the final step, Alice's detect 
$\sigma _A^{ij}$ in probability one, where \begin{equation}
					    \sigma_A^{ij} \stackrel{\rm
      def}{=} \frac{\sqrt{M_i \rho_A}N_j^{iT}\sqrt{\rho_A M_i}}{ \Tr \left
      (\sqrt{\rho_A}M_i \sqrt{\rho_A } N_j^{iT}
					   \right )}				   \end{equation} 
is Alice's
      state after Bob's measurement when a given state is $\ket{\Psi}$. 
Hence, $L^{ij}$ can be written down as $L^{ij}=\{ \sigma_A^{ij} >0 \}$.
\end{enumerate}
We define $\overline{S}_{\alpha, \leftrightarrow}(\ket{\Psi})$ as the optimal success probability under these two assumptions: 
\begin{align}\label{eq sec two-way def overline s}
 \quad& \overline{S}_{\alpha, \leftrightarrow}(\ket{\Psi})\nonumber \\
\stackrel{\rm def}{=}&  \max _{T} \Big \{\bra{\Psi}T\ket{\Psi} | \Tr T
 \le \alpha d_Ad_B, \nonumber \\
 \quad& T= \sum_{i \in I_0} \sum_{j=1}^{{\rm rank} M_i} \sqrt{M_i}\{ \sigma_A^{ij} >0 \}\sqrt{M_i}
 \otimes \ket{\xi_i^j}\bra{\xi_i^j}, \nonumber \\
\quad& 0 \le \sum _{i \in I_0} M_i \le I_A
\Big \}.
\end{align}
Then, $\overline{S}_{\alpha, \leftrightarrow}(\ket{\Psi})$ satisfies:
\begin{Lemma}
\begin{equation}
 S_{\alpha, \rightarrow}(\ket{\Psi})\le \overline{S}_{\alpha,
\leftrightarrow}(\ket{\Psi}) \le S_{\alpha, \leftrightarrow}(\ket{\Psi})
\end{equation}
\end{Lemma}
\begin{Proof}
The second inequality is trivial from the definition of
 $\overline{S}_{\alpha, \leftrightarrow}(\ket{\Psi})$. 
In order to see the first inequality, we need to choose 
$I_0 = \{ 1, \cdots, c \}$,
$M_i = \ket{i}\bra{i}$ for $1 \le i \le c-1$,
and $M_c = m_c \ket{c}\bra{c}$ in Eq.
(\ref{eq sec two-way def overline s}), 
where $c$ and $m_c$ are defined by
 Eq.(\ref{eq sec one-way locc def c}) and 
Eq.(\ref{eq sec one-way locc def m c}). Then, $T$ defined in Eq.
(\ref{eq sec two-way def overline s}) coincides the optimal one-way LOCC
 POVM given in Eq.(\ref{eq sec main def optimal one-way locc povm}) 
\hfill $\square$
\end{Proof}

The optimization of  
$\overline{S}_{\alpha, \leftrightarrow}(\ket{\Psi})$ can be
reduced as follows:
\begin{Lemma}
 \begin{align} \label{eq sec 2way max mik i in mathcal p dA}
  \quad& \overline{S}_{\alpha, \leftrightarrow}(\ket{\Psi})\nonumber\\ 
=& \max _{ \{ m_{i}^k \}_{i \in \mathcal{P}(d_A),  k \in i}}
\Big \{ \sum _{i, k} \lambda_k m_i^k  \  \Big | \  0 \le m_i^k, \sum_{i \in \mathcal{P}(d_A)}
m_i^k \le 1,\nonumber \\
\quad& \qquad  \sum _{i \in \mathcal{P}(d_A)} |i|\cdot
\frac{\sum _{k \in i}\lambda_k (m_i^k)^{2}}{\sum _{k \in i}\lambda_km_i^k }\le \alpha
d_Ad_B \Big  \}, 
 \end{align}
where $\mathcal{P}(d_A)$ is a power set (a set of all subsets) of 
$\{1, \cdots, d_A  \}$,
$|i|$ is a number of elements in the set $i$, 
and $\rho_A = \sum      _{k=1}^{d_A}\lambda_k \ket{k}\bra{k}$.
\end{Lemma}
\begin{Proof}
First, by straightforward calculations, we derive 
\begin{align}
 \bra{\Psi}T\ket{\Psi}=&\sum _i \rho_A M_i, \nonumber \\
\Tr T =& \sum _i {\rm rank} M_i \frac{\Tr \rho_AM_i^2}{\Tr \rho_AM_i}.
\nonumber
\end{align}
By using a pinching technique \cite{H06} in the eigen basis of $\rho_A$, we
      observe that $M_i$ can be chosen as to be a diagonal matrix in
      this basis. Moreover, we only need to consider 
POVM $\{ M_i \}_{i \in I_0}$ in which support of $M_i$ is different
      from $M_j$ for all $i \neq j$. This can be shown as follows:
Suppose $M_i$ and $M_j$ have the same support for an optimal $\{ M_i
      \}_{i \in I_0}$. 
We define a new POVM $T'$ by using
$\{ M_i' \}_{i \in I_0}$ which is defined as  $M_i'=M_i+M_j$, $M_j'=0$
      and $M_k'=M_k$ for all $k \neq i, j$. Then, we have 
\begin{align}
 \quad &  \left ( \Tr T - \Tr T' \right )/{\rm rank}M_i \nonumber\\
=&   
\Big ( \left < M_i \right >^2\left < M_j^2 \right >
-2\left < M_i M_j \right >\left < M_i \right >\left < M_j \right >
\nonumber \\ 
\quad & +\left < M_i^2 \right >\left < M_j \right >^2 \Big )/
\left ( \left < M_i \right >\left < M_j \right >\left < M_i+M_j \right >
\right ) \nonumber \\
\ge& \frac{\left (\sqrt{\left < M_i^2 \right >}\left < M_j \right
 >-\sqrt{\left < M_j^2 \right >}\left < M_i \right >\right  )^2}
{\left < M_i \right >\left < M_j \right >\left < M_i+M_j \right >}
\nonumber \\
\ge& 0,
\end{align}
where $\left < M \right >$ is abbreviation of $\Tr \rho_A M$,
and we used the Schwarz inequality in the first inequality.
Thus, we have $\Tr T' \le \Tr T$, and $T'$ is also optimal when $T$ is
 optimal.
Thus, we can choose $\mathcal{P}(d_A)$ as $I_0$.
Finally, by just defining $m_i^k$ as 
$M_i = \sum _{k \in i} m_i^k \ket{k}\bra{k}$, we derive Eq.(\ref{eq sec 2way max mik i in mathcal p dA}). 
\hfill $\square$
\end{Proof}
By direct calculation, we can check the function 
$\sum _{i \in \mathcal{P}(d_A)} |i|\cdot
\frac{\sum _{k \in i}\lambda_k m_i^k}{\sum _{k \in i}\lambda_km_i^k }$ is a convex function.
Therefore, the optimization problem in Eq.(\ref{eq sec 2way max mik i in
mathcal p dA}) is a convex optimization.   
Thus, its local optimum is the
global optimum, and we can easily access the optimum by numerics at
least for a small dimensional system.

Up to now, we have presented a mathematically rigorous reduction of
$\overline{S}_{\alpha, \leftrightarrow}(\ket{\Psi})$ and 
derived Eq.(\ref{eq sec 2way max mik i in mathcal p dA}). 
On the other hand, although we do not have any proof, numerical
calculations strongly suggest that 
$\overline{S}_{\alpha,\leftrightarrow}(\ket{\Psi})$ further can be reduced in
the following way: By adding further restrictions onto Eq.(\ref{eq sec 2way
max mik i in mathcal p dA}) as
$M_i=0$ for all 
$i \in  \mathcal{P}(d_A)$ except 
$i=\{1\}, \{1,2\}, \cdots, \{1, \cdots,   d_A \}$, we define  
 $\widetilde{S}_{\alpha,
 \leftrightarrow}(\ket{\Psi})$ as
\begin{align}\label{eq sec 2way def overline S}
\quad& \widetilde{S}_{\alpha, \leftrightarrow}(\ket{\Psi}) \nonumber \\
=&\max _{ \{ m_{i}^k \}_{1\le k \le i \le d_A}}\Big \{ \sum _{i, k} \lambda_k
 m_i^k  \  \Big | \  0 \le m_i^k, \sum_{i = k}^{d_A} m_i^k \le 1, \nonumber \\
\quad& \qquad \sum _{i = 1}^{d_A} i\cdot
\frac{\sum _{k =1}^i \lambda_k (m_i^k)^{2}}{\sum _{k = 1}^i
 \lambda_km_i^k }\le \alpha d_Ad_B \Big  \},
\end{align}
This optimization problem is a convex optimization with just $O(d_A^2)$
      parameters.
Our numerical calculations strongly suggest 
$\widetilde{S}_{\alpha, \leftrightarrow}(\ket{\Psi})=\overline{S}_{\alpha,
\leftrightarrow}(\ket{\Psi})$.
Even if this equality is not true, we trivially have 
$\widetilde{S}_{\alpha,\leftrightarrow}(\ket{\Psi}) 
\le \overline{S}_{\alpha,\leftrightarrow}(\ket{\Psi})$.
Thus, $\widetilde{S}_{\alpha, \leftrightarrow}(\ket{\Psi})$ is also a
lower bound of $S_{\alpha, \leftrightarrow}(\ket{\Psi})$. 
We can define the optimal type 2 error under the three assumptions as
$\widetilde{\beta}_{\ket{\Psi}, \leftrightarrow}(\alpha)\stackrel{\rm
def}{=} 1-\widetilde{S}_{\alpha, \leftrightarrow}(\ket{\Psi})$. Then, we
have $\widetilde{\beta}_{\ket{\Psi}, \leftrightarrow}(\alpha) \ge
\beta_{\ket{\Psi}, \leftrightarrow}(\alpha)$. This completes the proof of
Theorem \ref{main theorem two-way}.

\section{Global Hypothesis testing with a composite alternative
 hypothesis }\label{sec global}
As a preparation for the next section, we treat a global hypothesis
testing having a composite alternative hypothesis in this section. 
As we will prove in the next section, this relatively simpler 
hypothesis testing is
actually equivalent to the local hypothesis testing under separable
POVM. 
The organization of the section is as follows:
We explain the problem settings and the relation between the global
      hypothesis testing and the local hypothesis testing under
      separable POVM in the subsection A.
Then, we reduce the global hypothesis testing with a composite
alternative hypothesis to a hypothesis testing with a simple alternative
hypothesis with an additional restriction on POVM in the subsection B.
Finally, we derive analytical solutions for the problem in the
subsection C.

\subsection{Preliminary for the section}\label{sec global subsec preliminary}
In the conventional (classical) statistical inference, a hypothesis
testing normally has a composite hypothesis (a hypothesis
consists of a set of probability distributions) in practical situations, 
and a hypothesis testing with simple null and alternative hypotheses is
usually treated in pure theoretical motivation, like the Neyman-Pearson
lemma, Stein's lemma, and Chernoff bound. On the other hand, in
quantum statistical inference, so far, just a very limited number of works
treat a hypothesis testing with a composite hypothesis \cite{H09, N00,B10}. 
In this section, we add
one example into this category. Our example consists of a simple alternative
hypothesis and a composite null hypothesis 
on a single partite Hilbert-space $\Hi$:
A null hypothesis is a composite hypothesis, ``{\it an unknown state is in a
set $\{ \ket{\phi _{\vec{k}}}\}_{\vec{k}\in \mathbb{Z}_2^{d}}$}'' defined
as
\begin{equation}  \label{eq def phi vec k}
 \ket{\phi_{\vec{k}}} \stackrel{\rm def}{=} \frac{1}{\sqrt{d_A}} \sum _i (-1)^{k_i}\ket{i},
\end{equation}
and an alternative hypothesis is a
simple hypothesis 
``{\it an unknown state is a pure state $\ket{\psi}$}''.
An optimal success probability $X_{\epsilon}(\ket{\psi})$ of this problem is given as 
\begin{align}
 X_{\epsilon}(\ket{\psi}) \stackrel{\rm def}{=}& \max _T \{ \bra{\psi}T\ket{\psi} | T \in \B
  (\Hi),  0 \le T \le I, \nonumber \\
 \quad & \quad \quad \forall \vec{k}\in
 \mathbb{Z}_2^{d}, \bra{\phi _{\vec{k}}}T\ket{\phi_{\vec{k}}}| \le
 \epsilon ^2 \}, 
\end{align}
where $d$ is the dimension of the Hilbert-space $\Hi$.
Here, we define $\epsilon$ so that $\epsilon ^2$ is an upper bound of the
type 1 error. As we can easily see, this problem possesses a nice group
symmetry; that is, our composite hypothesis is
generated from a single state $\ket{\phi_0}$
by a group action of {\it phase flipping}: $\ket{i} \rightarrow -\ket{i}$. 
Actually, we will use this property to derive an analytical formula of 
$X_{\epsilon}(\ket{\psi})$.

In the next section, we will prove that this optimal success probability
$X_{\epsilon}(\ket{\psi})$ is equal to the optimal success 
probability of the local hypothesis testing under separable POVM 
$S_{\alpha, sep}(\ket{\Psi})$ with just rescaling parameters:
\begin{equation}\label{eq sec sep x epsilon = s''}
 X_{\sqrt{\alpha d_B}}(\ket{\psi}) = S_{\alpha, sep}(\ket{\Psi}),
\end{equation} 
where $\ket{\psi}$ is defined as $\ket{\psi}=\sum_{i=1}^{d_A} \sqrt{\lambda _i}\ket{i}$ by using the Schmidt
coefficients $\{ \lambda _i \}_{i=1}^{d_A}$ of $\ket{\Psi}$.
The aim of this subsection is deriving an analytical formula for
$X_{\epsilon}(\ket{\psi})$ as a preparation to derive an analytical 
formula for $S_{\alpha, Sep}(\ket{\Psi})$ by proving the above equality 
in the next section.
Thus, {\it we only treat a real $\ket{\psi}$ in this subsection}; that is,
$\ket{\psi}$ satisfies $\braket{i}{\psi} \in \mathbb{R}$ for all $i$.
In this case, without losing generality, we can always assume $\braket{i}{\psi} \ge 0$ for all $i$ 
by changing appropriate states in the basis as $\ket{i} \longrightarrow -\ket{i}$.  
Moreover, by changing the label of the basis, without losing generality,
we can also assume $\braket{i}{\psi} \ge \braket{i+1}{\psi}$ for all
$i$.
In the following discussion, we always choose the standard basis of $\Hi
_A$ as above.

\subsection{Reduction to a hypothesis testing with a simple alternative
  hypothesis}\label{sec global subsec reduction}
In this subsection, we show that the above global hypothesis testing
with a composite null hypothesis can be reduced to a global
hypothesis testing with an additional restriction on POVM.

First, we observe that an optimal $T$ can be chosen as ${\rm rank}T_0=1$.
\begin{Lemma}
\begin{align}
\quad& X_{\epsilon}(\ket{\psi}) \nonumber \\
=& \max_T \{ \bra{\psi}T\ket{\psi}\  |\  T \in \B
  (\Hi ),  0 \le T \le I, T= {\rm Re} T,  \nonumber \\
\quad& \quad {\rm rank}T=1, \forall {\vec k} \in \mathbb{Z}_2^{d},
 \bra{\phi _{\vec  k}}T\ket{\phi _{\vec k}} \le
 \epsilon ^2  \}
\end{align}
\end{Lemma}
\begin{Proof}
 About the condition $T= {\rm Re} T$, we have already seen that this
 condition does not change the value of the optimization problem at
 the last of the previous subsection. 
Thus, here, we only  treat the condition ${\rm rank} T =1$.

When $X_{\epsilon}(\ket{\psi})=0$, we can always choose $T=0$, which satisfies ${\rm rank} T=1$.
Thus, we assume $X_{\epsilon}(\ket{\psi})>0$, and, hence, 
$\bra{\psi}T\ket{\psi}>0$ for an optimal $T$. 

Suppose $T_0$ is an optimal POVM and there exists a state $\ket{\psi ^{\perp}} \in {\rm Ran}T_0$
 satisfying $\braket{\psi ^{\perp}}{\psi}$. 
Then, from $\det T_0 >0$ and the continuity of 
$\det (T_0 - p \ket{\psi ^{\perp}}\bra{\psi ^{\perp}})$ with respect to $p$,
there exists a minimum $p>0$ satisfying  $\det (T_0 - p \ket{\psi
 ^{\perp}}\bra{\psi ^{\perp}})>0$, 
where a determinant is defined only on ${\rm Ran}T_0$.
We call this optimal $p$ as $p_0$.
Then, a positive operator $T'_0 \stackrel{\rm def}{=} T_0 - p_0 \ket{\psi ^{\perp}}\bra{\psi
 ^{\perp}}$ satisfies ${\rm Ran} T'_0 = {\rm Ran}T_0 -1$,
 $\bra{\psi}T'_0\ket{\psi}=\bra{\psi}T_0\ket{\psi}$
and $\bra{\phi _{\vec k}}T'_0\ket{\phi _{\vec k}} =\bra{\phi _{\vec
 k}}T_0\ket{\phi _{\vec k}}-p_0|\braket{\phi_{\vec k}}{\psi ^{\perp}}|
 \le \epsilon^2$.
Thus, $T'_0$ is an optimal POVM whose range does not include a state $\ket{\psi^{\perp}}$.

By repeating the above argument, we can conclude that there exists
an optimal POVM $T_0$ whose range does not include any state 
$\ket{\psi ^{\perp}}$ satisfying $\braket{\psi ^{\perp}}{\psi}=0$. 
This optimal POVM $T_0$ should satisfy ${\rm rank}T_0=1$. 
We will show this fact by contradiction.
Suppose ${\rm rank} T_0 \ge 2$ for this $T_0$. 
Then, there exist states $\ket{e_0}, \ket{e_1} \in {\rm Ran} T_0$ satisfying
$\braket{e_0}{\psi} \neq 0$, $\braket{e_0}{e_1}=0$. Then, we can write down these states as 
\begin{align*}
 \ket{e_0}=& \alpha _0 \ket{\psi} + \sqrt{1-|\alpha
  _0|^2}\ket{\psi^{\perp}_0} \\
 \ket{e_1}=& \alpha _1 \ket{\psi} + \sqrt{1-|\alpha
  _1|^2}\ket{\psi^{\perp}_1},
\end{align*}
where $\alpha _0 \neq 0$, and the states $\ket{\psi ^{\perp}_0}$ and
 $\ket{\psi ^{\perp}_1}$ satisfies $\braket{\psi^{\perp}_0}{\psi}=\braket{\psi^{\perp}_1}{\psi}=0$.
Then, we can conclude that an operator $\ket{\psi^{\perp'}}$ defined as 
\begin{align*}
\quad& \ket{\psi^{\perp'}} \nonumber \\
 \stackrel{\rm def}{=}& \alpha _1 \ket{e_0}-\alpha_0 \ket{e_1}
\\ =& \alpha_1 \sqrt{1- |\alpha
 _0|^2}\ket{\psi^{\perp}_0}-
\alpha_0 \sqrt{1- |\alpha _1|^2}\ket{\psi^{\perp}_1} 
\end{align*}
satisfies $\ket{\psi^{\perp'}} \neq 0$.
Since $\ket{\psi^{\perp'}} \in {\rm Ran}T_0$ and
 $\braket{\psi^{\perp'}}{\psi}=0$,
this is a contradiction.
\hfill $\square$ 
\end{Proof}

From the previous lemma, we can always choose an optimal POVM $T$ as
$T=\ket{\phi}\bra{\phi}$.
Moreover, from non-negativity of $\braket{i}{\psi}$, we can also assume $\braket{i}{\phi} \ge 0$ for all $i$ as follow:
\begin{Lemma}
\begin{align}
X_{\epsilon}(\ket{\psi}) =& \max_{\ket{\phi}} \big \{
 \braket{\psi}{\phi}^2 \ \big | \ 
\ket{\phi}\in \Hi,  \| \ket{\psi}
 \|^2 \le 1, \nonumber \\ 
\quad & \quad \forall i, \braket{i}{\psi} \ge 0, \forall \vec{k} \in \mathbb{Z}^{d}_2, |\braket{\phi
 _{\vec k}}{\phi}|^2 \le \epsilon ^2 \big \}.\nonumber  
\end{align}
\end{Lemma}
\begin{Proof}
First, we can always choose an optimal state $\ket{\phi}$ as
 $\braket{\psi}{\phi} \ge 0$.
We define coefficients $b_i$ as $\ket{\phi}=\sum _i b_i \ket{i}$.
Suppose there exists $i_0$ such that $b_{i_0} <0$ for an
 optimal $\ket{\phi}$  satisfying $\braket{\psi}{\phi} \ge 0$. 
We define $\ket{\phi'}=\sum _i b'_i\ket{i}$ as $b'_{i_0}=-b_{i_0}$
and $b'_i=b_i$ for all $i \neq i_0$.
Then, $\braket{\phi_{\vec k}}{\phi} \le \epsilon^2$ for all $\vec k$ guarantees
$\braket{\phi _{\vec k}}{\phi'} \le \epsilon^2$, 
and $\braket{i}{\psi} \ge 0$ for all $i$ guarantees
 $\braket{\psi}{\phi'}>\braket{\psi}{\phi}$.
This is a contradiction. 
Therefore, an optimal $\ket{\phi}$ satisfying $\braket{\psi}{\phi} \ge 0 $
must satisfy $\braket{i}{\phi}\ge 0$ for all $i$.
In other words, we can always choose an optimal state $\ket{\psi}$ as
 above.
\hfill $\square$
\end{Proof}

Finally, we can transform $X_{\epsilon}(\ket{\psi})$
in the following form:
\begin{Lemma}
\begin{align}\label{eq sep x epsilon = max last}
  X_{\epsilon}(\ket{\psi}) =& \Big [ \max \big \{ 
\braket{\psi}{\phi} \ \big | \ \ket{\phi}\in \Hi, \| \ket{\phi}\|^2\ge 1, \nonumber \\
\quad & \qquad \forall i, \braket{i}{\phi}\ge \braket{i+1}{\phi} \ge 0,
 \braket{\phi_d}{\phi}\le \epsilon   \big \} \Big ]^2,
 \end{align}
where $\ket{\phi _{j}}$ is defined as 
\begin{equation} \label{eq sec sep def phi_d}
\ket{\phi _{j}}=  \frac{1}{\sqrt{j}}\sum _{i=1}^j \ket{i}.
\end{equation}
\end{Lemma}
\begin{Proof}
From the previous lemma, we can always choose an optimal state
 $\ket{\phi}=\sum _i b_i \ket{i}$ as one satisfying $b_i \ge 0$ for all $i$.
Suppose there exists a pair $i_0 < i_1$ such that  $b_{i_0}< b_{i_1}$.
We define $\ket{\phi'}=\sum _i b'_i \ket{i}$ as $b'_{i_0}=b_{i_1}$,
 $b'_{i_1}=b_{i_0}$, and $b'_i=b_i$ for all $i \neq i_0, i_1$. 
 Then, $\ket{\phi'}$ satisfies 
$|\braket{\phi _{\vec k}}{\phi'}|^2 \ge \epsilon^2$ for all
${\vec k}\in \mathbb{Z}_2^d $ 
and $\braket{\psi}{\phi'}>\braket{\psi}{\phi}$. 
Thus, $\ket{\psi}$ is not an optimal state; this is a contradiction.
Therefore, an optimal state $\ket{\phi}$ with $b_i \ge 0$ satisfies $b_i \ge b_{i+1}$ for all $i$.
This optimal state apparently satisfies 
\begin{equation}
 \braket{\phi _d}{\phi}^2 \ge |\braket{\phi _{\vec k}}{\phi}|^2.
\end{equation}
Thus, we can replace the condition $|\braket{\phi _{\vec k}}{\phi}|^2 \le
 \epsilon ^2$ by the condition $\braket{\phi _d}{\phi} \le \epsilon$ for
 this optimal state.
\hfill $\square$
\end{Proof}

The optimization problem in Eq.(\ref{eq sep x epsilon = max last}) does
not contain a composite hypothesis, but is a hypothesis testing 
of two simple hypotheses $\ket{\psi}$ and $\ket{\phi _d}$ with
an additional restriction on the form of the POVM.
In the next subsection, we analytically solve this optimization problem.  

\subsection{Derivation of analytical solutions of $X_{\epsilon}(\ket{\psi})$}\label{sec global subsec solution}
First, we give solutions of $X_{\epsilon}(\ket{\psi})$ for two trivial cases.
When $\ket{\psi}=\ket{\phi_d}$, we can easily see $X_{\epsilon}(\ket{\phi_d})=\epsilon$.
When $\braket{\psi}{\phi_d} \le \epsilon$, we can choose an optimal
vector $\ket{\phi}$ as $\ket{\phi}=\ket{\psi}$.
Hence, $X_{\epsilon}(\ket{\phi_d})=1$.

For $\ket{\psi}\neq \ket{\phi _d}$, we derive the following lemma:
\begin{Lemma} \label{lemma sec sep x epsilon enumerate}
 Suppose $d \ge 2$, $\ket{\psi}\neq \ket{\phi_d}$,
and $\ket{\phi}$ attains the optimal of  Eq.(\ref{eq sep x epsilon = max
 last}).
Then, at least, one of the following two statement is true:
\begin{enumerate}
 \item $\ket{\phi}\in {\rm span} \left \{\ket{\psi}, \ket{\phi _d}
       \right \}$, $\| \ket{\phi} \|=1$.
 \item There exists an optimal state $\ket{\phi'}$  satisfying $\braket{d}{\phi'}=0$.
\end{enumerate}
\end{Lemma}
\begin{Proof}
 Suppose $\ket{\phi}$ attains the optimal of Eq.(\ref{eq sep x epsilon = max
 last}). 
First, we can uniquely decompose $\ket{\phi}$ as follows:
\begin{equation}
 \ket{\phi}=\alpha \ket{\psi}+\beta\ket{\phi _d}+\ket{y},
\end{equation}
where $\alpha$ and $\beta$ are real numbers, and $\ket{y}$ satisfies
 $\ket{y} \perp \ket{\phi _d}$ and $\ket{y} \perp \ket{\psi}$.
Then, we define a Schmidt orthogonalized state $\ket{\phi _d^{\perp}}$ on a
 subspace ${\rm span} \{\ket{\psi}, \ket{\phi _d} \} $ as 
\begin{equation}\label{eq def of phi d perp}
 \ket{\phi _d^{\perp}} \stackrel{\rm def}{=} \frac{\ket{\psi} -
  c \ket{\phi_d}}{\left \| \ket{\psi} -
  c \ket{\phi_d}  \right \|},
\end{equation}
where $c \stackrel{\rm def}{=} \braket{\psi}{\phi_d}$. 
By the definition, $\ket{\phi _d^{\perp}}$ satisfies $\braket{\phi _d^{\perp}}{\psi}>0$.
Also, by defining the coefficients $\{ \xi _i \}_{i=1}^d$ as
 $\ket{\phi _d ^{\perp}}=\sum _i \xi _i \ket{i}$, these coefficients
 satisfy $\xi _i \ge \xi _{i+1}$ for all $i$. Moreover, the fact
 $\braket{\phi_d^{\perp}}{\phi _d}=0$ 
guarantees that there exists a natural number $l$ satisfying 
$\xi _l \ge 0 > \xi _{l+1} $.

First, we consider the case when $\| \ket{\phi} \| <1$.
In this case, actually, $\ket{\phi}$ satisfies $\braket{d}{\phi}=0$.
 This fact can be proven  by contradiction as follows:
Suppose $\braket{d}{\phi}>0$.
Then, we can choose a small number $\delta >0$
such that a vector $\ket{\phi'} \stackrel{\rm def}{=} \ket{\phi}+\delta \ket{\phi
 _d^{\perp}}$ 
satisfies  $\| \ket{\phi '}\| \le 1$, 
$\braket{\phi _d}{\phi'}=\braket{\phi _d}{\phi}$,
$\braket{i}{\phi'} \ge 0$ for all $i$, 
and $\braket{\psi}{\phi'}=\braket{\psi}{\phi} + \alpha
 \braket{\psi}{\phi _d^{\perp}}> \braket{\psi}{\phi}$.
Thus, $\ket{\phi}$ is not an optimal state.
This is a contradiction.
Hence, in this case, a state $\ket{\phi}$ itself is an optimal state
 satisfying 
the condition $\braket{d}{\phi}=0$.

Second, we consider the case $\| \ket{\phi} \| =1$.
In this case, we consider the case $\alpha <0$ and the case $\alpha \ge
 0$ separately.
Thus, we consider the case when $\| \ket{\phi} \|=1$ and $\alpha <0$.
In this case, $\braket{d}{\phi}=0$ is proven by contradiction as follows:
Suppose $\braket{d}{\phi}=0$.
Then, we can choose a real number $\gamma$ 
so that it satisfies 
\begin{equation}\label{eq 0 le gamma le -alpha}
0 \le \gamma \le -\alpha \| \ket{\psi} - c
 \ket{\phi_d}\|, 
\end{equation}
and a vector $\ket{\phi'} \stackrel{\rm def}{=} \gamma \ket{\phi
 _d^{\perp}}+\ket{\phi}$  satisfies 
$\braket{d}{\phi'} \ge 0$.
This vector $\ket{\phi'}$ satisfies
 $\braket{\phi_d}{\phi'}=\braket{\phi_d}{\phi}$,
$\braket{i}{\phi'}\ge \braket{i+1}{\phi'} \ge 0$ for all $i$,
and $\braket{\psi}{\phi'}= \gamma \braket{\psi}{\phi
 _d^{\perp}}+\braket{\psi}{\phi} > \braket{\psi}{\phi}$.
Furthermore, from Eq.(\ref{eq 0 le gamma le -alpha})
and $\| \ket{\phi}\|=1$, 
the formula 
\begin{equation}
 \ket{\phi'}= \left ( c+\beta \right ) \ket{\phi_d}+ \left (\alpha \| \ket{\psi} - c\ket{\phi_d}
				    \|+\gamma \right )\ket{\phi
				    _d^{\perp}} + \ket{y}
\end{equation}
guarantees $\| \ket{\phi'}\| <1$.
Thus, $\ket{\phi}$ is not an optimal. 
This is a contradiction.
Therefore, we have $\braket{d}{\phi}=0$ in this case.

Next, we consider the case when  $\| \ket{\phi} \|=1$ and $\alpha \ge 0$.
We define a vector $\ket{x}$ as $\ket{x} \stackrel{\rm def}{=} \alpha
 \ket{\psi} +\beta \ket{\phi _d }$, and its coefficients 
$\{ x_i \}_{i=1}^d$ as $\ket{x} = \sum _{i=1}^d x_i \ket{i}$,
which apparently satisfy $x_i \ge x_{i+1}$ for all $i$.
In this case, we also consider the cases $x_d =0$,
$x_d >0$, and $x_d < 0$ separately.

First, we consider the case $x_d =0$.
In this case, this vector $\ket{x}$ is apparently an optimal
 vector satisfying $\braket{d}{x}=0$. 

Second, we consider the case $x_d > 0$. 
In this case, this vector $\ket{x}$ is apparently an optimal
 vector satisfying 
$\ket{x} \in {\rm span} \left \{ \ket{\psi}, \ket{\phi_d}\right \}$. 
Moreover, we can prove that this vector also satisfies $\| \ket{x}\|=1$
 by contradiction as follows: 
Suppose $\| \ket{x}\|<1$. 
Then, there exists a small number $\delta >0$ such that
a new vector $\ket{x'} \stackrel{\rm def}{=} \alpha\ket{\psi}+ \beta
 \ket{\phi_d} + \delta \ket{\phi_d^{\perp}}$ satisfying 
$\| \ket{x'}\| \le 1$ and $\braket{d}{x'} \ge 0$. 
This vector $\ket{x'}$ satisfies 
$\braket{\phi _d}{x'}=\braket{\phi _d}{\phi}$, 
$\braket{i}{x'} \ge \braket{i+1}{x'} \ge 0$ for all $i$,
and $\braket{\psi}{x'}=\braket{\psi}{\phi}+\delta \braket{\psi}{\phi
 _d^{\perp}} > \braket{\psi}{\phi}$.
Thus, $\ket{\phi}$ is not an optimal vector.
This is contradiction.
Therefore, $\| \ket{x} \|=1$, 
and this means $\ket{x}=\ket{\phi}$.
Hence, $\ket{\phi} \in {\rm span}\left \{ \ket{\psi}, \ket{\phi _d}
 \right \}$.

Finally, we consider the case when $x_d <0$.
In this case, there exists a natural number $m \le d+1$ such that 
$x_m \ge 0 > x_{m+1}$. Here, we define a one-parameter family of vectors
 $\{ \ket{z _{\delta}} \}_{ 0 \le \delta \le 1}$ as 
$\ket{z_{\delta}} \stackrel{\rm def}{=} \ket{x} + \delta \ket{y}$.
Hence, $\ket{z_0}=\ket{x}$ and $\ket{z_1}=\ket{\phi}$.
For all $0 \le \delta \le 1$, this family satisfies
 $\braket{\psi}{z_{\delta}}=\braket{\psi}{x}=\braket{\psi}{\phi}$,
$\braket{\phi_d}{z_{\delta}}=\braket{\phi_d}{x}=\braket{\phi_d}{\phi}$,
and $\| \ket{z_{\delta}} \| \le \| \ket{\phi} \|=1$.
We define a function $f(\delta)$ as 
$f(\delta)= \left ( z_{\delta,1}, \cdots, z_{\delta, d} \right )$,
where $z_{\delta,i}$ is defined as $\ket{z_{\delta}}=\sum _{i=1}^d z_{\delta,i}\ket{i}$.
Then, the point $\left ( z_{1,1}, \cdots, z_{1, d} \right )$
 satisfies $z_{1,i} \ge 0$ for all $i$,
and the point $\left ( z_{0,1}, \cdots, z_{0, d} \right )$ satisfies
 $z_{0,i} \ge 0$ for $i \le m$ and $z_{0,i}<0$ for $i \ge m+1$.
Hence, a connecting curve $f(\delta)$ on the $d$-dimensional space 
starts from the outside and goes into the region $\{ x_i \ge 0, \forall i\}$. 
Therefore, this curve must across the boundary of this region in
 somewhere between the start point $\delta=0$ and the end point $\delta=1$.
Thus, there exits $0 < \delta_0 \le 1$ such that $z_{\delta_0,i} \ge 0$
 for all $i$ and there exits $i_0$ satisfying $z_{\delta_0,i_0}=0$.
Next, we define new coefficients $\{ z'_i \}_{i=1}^d$ which are derived
 by reordering $\{ z_{\delta_0, i} \}_{i=1}^d$
so that they satisfy $z'_i \ge z'_{i+1} \ge 0$.  
Then, a state $\ket{\phi'}$ defined as 
$\ket{\phi'}\stackrel{\rm def}{=}\sum_i z'_i \ket{i}$ 
satisfies
$\braket{\psi}{\phi'}\ge\braket{\psi}{\phi}$,
 $\braket{\phi_d}{\phi'}=\braket{\phi_d}{\phi}$,
$\| \ket{\phi'}\| \le \| \ket{\phi} \|$, $\braket{i}{\phi'} \ge 0$.
Therefore, $\ket{\phi'}$ is actually an optimal state satisfying $\braket{d}{\phi'}=0$.
\hfill $\square$
\end{Proof}

The following lemma gives a non-trivial solution of the optimization
problem.
\begin{Lemma} \label{lemma sec sep x epsilon special solution}
Consider the case when $d \ge 2$ and 
$\epsilon < \braket{\phi_d}{\psi} <1$.
Suppose $\ket{\phi}$ defined as 
\begin{eqnarray}\label{eq def optimal phi}
 \ket{\phi} \stackrel{\rm def}{=} \frac{\sqrt{1-\epsilon ^2}\ket{\psi} -
  \left ( c\sqrt{1-\epsilon ^2} - \epsilon \sqrt{1-c^2} \right )\ket{\phi _d}}{\sqrt{1-c^2}},
\end{eqnarray}
where $c \stackrel{\rm def}{=} \braket{\psi}{\phi _d}$,
satisfies $\braket{d}{\phi} \ge 0$.
Then, $\ket{\phi}$ attains the optimum of Eq.(\ref{eq sep x epsilon = max last}).
\end{Lemma} 
\begin{Proof}
We define a new function $X'_{\epsilon}(\ket{\psi})$ as follows:
\begin{align}\label{eq def x' epsilon}
\quad & X'_{\epsilon}(\ket{\psi}) \nonumber \\ 
=& \max_{\ket{\phi}} \big \{
 \braket{\psi}{\phi}^2 \ \big | \ 
\ket{\phi}\in \Hi,  \| \ket{\psi}
 \|^2 \le 1, |\braket{\phi
 _{d}}{\phi}|^2 \le \epsilon ^2 \big \}.\nonumber  
\end{align}
Then, by the definition, $X'_{\epsilon}(\ket{\psi})$ satisfies $X_{\epsilon}(\ket{\psi}) \le X'_{\epsilon}(\ket{\psi})$.
Thus, if $\ket{\phi}$ defined by Eq.(\ref{eq def optimal phi}) attains
 the optimum of $X'_{\epsilon}(\ket{\psi})$, and also satisfies
 $\braket{d}{\phi} \ge 0$, this vector $\ket{\phi}$  apparently also attains
 the optimum of  $X_{\epsilon}(\ket{\psi})$.
In the remaining part of this proof, we prove actually this is the case;
this $\ket{\phi}$ is an optimal vector for $X'_{\epsilon}(\ket{\psi})$.

Suppose $\ket{\phi}$ is an optimal vector of Eq.(\ref{eq def x' epsilon}).
Then, by the definition of $X'_{\epsilon}(\ket{\psi})$, 
$\ket{\phi}$ is apparently on the subspace 
${\rm span}\left \{ \ket{\psi}, \ket{\phi _d} \right \}$.
Thus, we define $r$, $\theta$ and $\xi$ satisfying $r>0$, $-\pi \le \theta \le \pi $ and 
$-\pi \le \xi \le \pi $, respectively as 
\begin{eqnarray}\label{eq decomp psi phi 2}
 \ket{\psi} &=& \cos \theta \ket{\phi _d} + \sin \theta \ket{\phi
  _d^{\perp}} \nonumber \\
\ket{\phi} &=& r \left ( \cos \xi \ket{\phi _d} + \sin \xi \ket{\phi
  _d^{\perp}}\right ) , 
\end{eqnarray}
where $\ket{\phi_d^{\perp}}$ is defined by Eq. (\ref{eq def of phi d
 perp}).
By the definitions, we have $\cos \theta = \braket{\psi}{\phi_d}$, 
$\sin \theta = \braket{\psi}{\phi_d^{\perp}}$, 
$\cos \xi = \braket{\psi}{\phi_d}/r$, and $\sin \xi = \braket{\psi}{\phi_d^{\perp}}/r$.
Thus, $\braket{\psi}{\phi_d} > 0$ and
 $\braket{\psi}{\phi_d^{\perp}}>0$ guarantee $0 < \theta <\frac{\pi}{2}$.
$\braket{\psi}{\phi} >0$ guarantees 
$ - \frac{\pi}{2} \le  \theta - \xi \le \frac{\pi}{2}$.

First, we prove $\xi \ge 0$ by contradiction.
Suppose $\xi <0$; that is, $-\frac{\pi}{2} \le \xi <0$.
Then, defining  $\ket{\phi'}$ by using $\xi' \stackrel{\rm def}{=} -\xi$ instead of $\xi$ in
  Eq.(\ref{eq decomp psi phi 2}), we have $|\braket{\phi_d}{\phi'}| \le
 \epsilon$, $\| \ket{\phi'} \| = \| \ket{\phi}\| \le 1$.
Moreover, the inequalities $|\xi'-\theta| < |\xi'| +|\theta| = -\xi
 +\theta \le \frac{\pi}{2}$ guarantee
 $\braket{\psi}{\phi'}>\braket{\psi}{\phi}$.
Thus, $\ket{\phi}$ is not optimal; this is contradiction.
Therefore, $\xi$ satisfies $\xi \le 0$.
 
Second, we prove $r=1$ by contradiction.
Suppose $r<1$.
Then, we can choose a small number $\delta >0$ 
such that a state $\ket{\phi'}$ defined as 
$\ket{\phi'} \stackrel{\rm def}{=} \ket{\phi} + \delta
 \ket{\phi_d^{\perp}}$ satisfies $\| \ket{\phi'}\|\le 1$.
Then, this state $\ket{\phi'}$ satisfies
$|\braket{\phi_d}{\phi'}|=|\braket{\phi_d}{\phi}|\le \epsilon$,
and $\braket{\psi}{\phi'}>\braket{\psi}{\phi}$.
Thus, $\ket{\phi}$ is not optimal; this is contradiction.
Therefore, $r$ satisfies $r=1$.

Finally, we prove  $\epsilon = \braket{\phi_d}{\phi}$ by contradiction.
Suppose $\epsilon > \braket{\phi_d}{\phi}=\cos \xi$.
In this case, we can choose a small number $\delta >0$ such that 
$\xi' \stackrel{\rm def}{=}\xi -\delta$ satisfies $|\cos \xi'| \le
 \epsilon$.
Then, defining $\ket{\phi'}$ by using $\xi'$ and $r=1$ in Eq.(\ref{eq
 decomp psi phi 2}),
we derive $\braket{\psi}{\phi'}=\cos (\theta - \xi') > \cos (\theta -\xi)=\braket{\psi}{\phi}$.
Thus, $\ket{\phi}$ is not optimal; this is contradiction.
Therefore, an optimal vector $\ket{\phi}$ is the unique vector satisfying
 $\epsilon=\braket{\phi_d}{\phi}$ and $0 \le \xi \le
 \frac{\pi}{2}+\theta \le \pi$.
Eq.(\ref{eq decomp psi phi 2}) guarantees that 
this vector $\ket{\phi}$ can be written in the form of Eq.(\ref{eq def
 optimal phi}).

\hfill $\square$
\end{Proof}

At the next step, by means of Lemma \ref{lemma sec sep x epsilon
enumerate}
and Lemma \ref{lemma sec sep x epsilon special solution}
we derive the following lemma:
\begin{Lemma}\label{lemma sec sep zennkashiki}
Suppose $d \ge 2$ and $ \epsilon < \braket{\phi _d}{\psi} < 1$.
Define $\ket{\phi}$ as Eq.(\ref{eq def optimal phi}).
Then, when $\braket{d}{\phi} \ge 0$, $\ket{\phi}$ is an optimal vector
 for Eq.(\ref{eq sep x epsilon = max last}),
and when $\braket{d}{\phi}<0$, there exists an optimal vector
 $\ket{\phi'}$ satisfying $\braket{d}{\phi'}=0$ for Eq.(\ref{eq sep x epsilon = max
 last}).
\end{Lemma}
\begin{Proof}
We first consider the case where the optimal vector $\ket{\phi}$
 satisfies  $\ket{\phi} \in
 {\rm span} \left \{ \ket{\psi}, \ket{\phi _d} \right \}$ and $\| \ket{\phi}\|$=1.
In this case, we can define notations as Eq(\ref{eq decomp psi phi 2}) in the last section again.
Here, we choose $\theta$ and $\xi$ to be $-\pi < \theta \le \pi$ and $ -\pi < \xi \le \pi$ for convenience.
By the definitions, we again have $0 < \theta <\frac{\pi}{2}$ and $ - \frac{\pi}{2} \le  \theta - \xi \le \frac{\pi}{2}$.
In the similar way, $\braket{\phi}{\phi_d} >0$ guarantees $-\frac{\pi}{2} \le \xi \le
 \frac{\pi}{2}$ in this case.

First, we prove $\xi > 0$ by contradiction.
Suppose $\xi \le 0$. 
As we explained in the proof of  Lemma \ref{lemma sec sep x epsilon
enumerate}, 
there exist a natural number $l \le d-1$ such that 
$\braket{l+1}{\phi_d^{\perp}} < 0$.
From this fact and Eq.(\ref{eq decomp psi phi 2}),
$\xi \le 0$ guarantees $\braket{l+1}{\phi}<0$.
Thus, $\ket{\phi}$ is not an optimal vector of Eq.(\ref{eq sep x epsilon = max
 last}). This is contradiction.
Therefore, $\xi$ satisfies  $0< \xi \le \frac{\pi}{2}$.

Second, we can prove $\theta < \xi$ in the completely same discussion
 of the previous lemma.
Therefore, $\theta$ and $\xi$ satisfy $0 < \theta < \xi \le
 \frac{\pi}{2}$.
When $\ket{\psi}$ satisfies $\braket{\phi}{\phi_d}=\epsilon$,
$\ket{\psi}$ can be written down as Eq.(\ref{eq def optimal phi}).
Thus, since we are assuming the optimality of $\ket{\psi}$, 
$\ket{\psi}$ satisfies $\braket{d}{\psi} \ge 0$.
On the other hand, when $\ket{\psi}$ satisfies $\braket{\phi}{\phi_d} <
 \epsilon$,
we can prove $\braket{d}{\phi}=0$ by contradiction.
Suppose $\braket{d}{\phi}>0$ and $\braket{\phi}{\phi_d}< \epsilon$.
Then, we can choose a small number $\delta$ such that
a vector $\ket{\phi'}\stackrel{\rm def}{=}\cos (\xi - \delta)\ket{\phi_d}+ \sin (\xi
 -\delta)\ket{\phi_d^{\perp}}$ satisfies 
$\braket{\phi_d}{\phi'} \le \epsilon$, and
$\braket{i}{\phi'} \ge 0$ for all $i$.
Since $\ket{\phi'}$ satisfies
 $\braket{\psi}{\phi'}>\braket{\psi}{\phi'}$,
$\ket{\phi}$ is not optimal; this is contradiction.
Therefore, when $\braket{\phi_d}{\phi} < \epsilon$, 
$\ket{\phi}$ satisfies $\braket{d}{\phi} = 0$.

Now, we consider the case there is no agreement whether an optimal
 vector $\ket{\phi}$ satisfies
$\ket{\phi} \in  {\rm span }\{ \ket{\psi}, \ket{\phi _d}\}$, or not.
When $\ket{\phi}$ defined by  Eq.(\ref{eq def optimal phi}) satisfies
$\braket{d}{\phi}\ge 0$, then from Lemma  \ref{lemma sec sep x epsilon
 special solution},
this vector $\ket{\phi}$ is an optimal vector.
Then, we consider the case when $\ket{\phi}$ defined by  Eq.(\ref{eq def optimal phi}) does
 not satisfy $\braket{d}{\phi}\ge 0$.
In this case, if there exists an optimal vector $\ket{\phi}$ on  
the subspace ${\rm span} \{ \ket{\psi}, \ket{\phi _d}\}$ satisfying  $\| \ket{\psi} \|=1$, 
$\ket{\phi}$ should satisfy $\braket{\phi}{\phi_d} < \epsilon$,
and thus, $\braket{d}{\phi}=0$ from the above discussion.
Otherwise, there is no optimal vector satisfying $\| \ket{\phi} \|=1$ on
the  subspace ${\rm span} \{ \ket{\psi}, \ket{\phi _d}\}$.
In this case, from Lemma \ref{lemma sec sep x epsilon 
enumerate},
there exists an optimal vector $\ket{\phi'}$ satisfying $\braket{d}{\phi'}=0$.
Therefore, the statement of the present lemma is true.
\hfill $\square$
\end{Proof}

Finally, from the above lemma, we derive the following theorem,
which gives a complete analytical formula for the
optimal success probability $X_{\epsilon}(\ket{\psi})$ and the optimal 
strategy of the global
hypothesis testing considering in this section:
\begin{Theorem}\label{theorem main sec sep 1}
Suppose $\ket{\psi}\in \Hi$ can be written down as 
$\ket{\psi}=\sum _i \sqrt{\lambda_i} \ket{i}$.
 Define a natural number $\eta$ as 
\begin{align}
\quad & \eta \nonumber \\
\stackrel{\rm def}{=}& \min _{l \in \mathbb{N}} \Big \{ l \  \Big
 | \ l \le d,
 \epsilon _l < \braket{\phi_l}{\psi_l},  \ket{\psi _l} \neq  \ket{\phi
 _l}, \braket{l}{\phi'_l} <0 \Big \} -1.
\end{align}
In the above formula, $\epsilon_l$ is defined as $\epsilon _l \stackrel{\rm def}{=}
 \sqrt{\frac{d}{l}}\epsilon$,
a state $\ket{\psi_l}$ is defined as 
\begin{equation}\label{eq sec sep def ket psi_l}
 \ket{\psi_l} \stackrel{\rm def}{=}
\sum _{i=1}^l \sqrt{\lambda_i}\ket{i}/\sqrt{\sum_{i=1}^l \lambda_i},
\end{equation} 
a state
 $\ket{\phi_l}$ is defined by Eq.(\ref{eq sec sep def phi_d}), and a
 state $\ket{\phi'_l}$ is defined as
\begin{equation} \label{eq sec sep theorem 1 optimal state non-trivial}
 \ket{\phi'_l}=\frac{\sqrt{1-\epsilon _{l}^2}\ket{\psi_{l}}- \left (
c_{l}\sqrt{1-\epsilon_{l}^2}-\epsilon_{l} \sqrt{1-c_{l}^2} \right )\ket{\phi_{l}}}{\sqrt{1-c_{l}^2}},
\end{equation}
where $c_l$ is defined as $c_l\stackrel{\rm def}{=} \braket{\psi_l}{\phi_l}$.
Then, $\eta$ satisfies $\eta \ge 1$, and the following statements are true:
\begin{enumerate}
 \item In the case when $\epsilon _{\eta} \ge       \braket{\phi_{\eta}}{\psi_{\eta}}$, 
\begin{equation}\label{eq sec sep theorem 1 solution trivial}
X_{\epsilon}(\ket{\psi})=\sum _{i=1}^{\eta} \lambda_i,
\end{equation}
and a state $\ket{\phi}=\ket{\psi_{\eta}}$ attains the optimum.

\item In the case when $\epsilon_{\eta} <    \braket{\phi_{\eta}}{\psi_{\eta}}$,
$X_{\epsilon}(\ket{\psi})$ is given as
\begin{equation}\label{eq sec sep theorem 1 solution non-trivial}
 X_{\epsilon}(\ket{\psi})=  \left (\sum_{i=1}^{\eta}\lambda_i \right)\cdot 
\left ( \sqrt{1-\epsilon_{\eta}^2} \sqrt{1-c_{\eta}^2} +
 \epsilon_{\eta} c_{\eta} \right )^2.
\end{equation}
A vector $\ket{\phi}$ attaining this optimum is given 
as $\ket{\phi}=\epsilon _{\eta} \ket{\phi_{\eta}}$ in the case
      $\ket{\psi_{\eta}}=\ket{\phi_{\eta}}$,
and $\ket{\phi}=\ket{\phi'_{\eta}}$
in the case $\ket{\psi_{\eta}} \neq \ket{\phi _{\eta}}$, respectively
\end{enumerate}
\end{Theorem}
Here, we add one remark: When $\eta \ge 2$, $\epsilon_{\eta} <
\braket{\phi_{\eta}}{\psi_{\eta}}$ and 
$\ket{\psi_{\eta}}=\ket{\phi_{\eta}}$,
by redefining $\ket{\phi}=\epsilon_{\eta}
\ket{\phi_{\eta}}+\sqrt{1-\epsilon_{\eta}^2}\ket{\phi_{\eta}^{\perp}}$,
we can make $\ket{\phi}$ be a {\it normalized vector}. 
Therefore, in the case $\eta \ge 2$, we can always choose $\ket{\phi}$ as a
normalized vector; that is, $T$ is a pure state.  

\begin{Proof}
Suppose the formula
\begin{equation} \label{eq sec sep x epsilon zenkashiki}
 X_{\epsilon}(\ket{\psi})=\left ( \sum_{i=1}^{l}\lambda_i \right )\cdot X_{\epsilon _{l}}(\ket{\psi_{l}})
\end{equation} 
holds for $l=\eta$.
Then, in the case $\braket{\psi_{\eta}}{\phi_{\eta}} \le \epsilon_{\eta}$, since
$\ket{\phi}={\psi_{\eta}}$ attains $X_{\epsilon_{\eta}}(\ket{\psi_{\eta}})=1$, we derive
 Eq.(\ref{eq sec sep theorem 1 solution trivial}).
In the case $\braket{\psi_{\eta}}{\phi_{\eta}} > \epsilon_{\eta}$,
from the definition of $\eta$,
either $\ket{\psi_{\eta}}=\ket{\phi_{\eta}}$ or 
$\braket{\eta}{\phi'_{\eta}} \ge 0$ holds. When  $\ket{\psi_{\eta}}=\ket{\phi_{\eta}}$, 
a state $\ket{\phi}=\epsilon_{\eta} \ket{\phi_{\eta}}$ attains 
the optimum $X_{\epsilon_{\eta}}(\ket{\phi_{\eta}})=\epsilon_{\eta}$.
Thus, Eq.(\ref{eq sec sep theorem 1 solution non-trivial}) holds.
When $\braket{\eta}{\phi'_{\eta}} \ge 0$ holds,
from Lemma \ref{lemma sec sep x epsilon special solution}, 
a state $\ket{\phi'_{\eta}}$ given by Eq.(\ref{eq sec sep theorem 1 optimal state non-trivial})
attains the optimum and $X_{\epsilon}(\ket{\psi})$ is given by
 Eq.(\ref{eq sec sep theorem 1 solution non-trivial}).
Hence, all the statements hold under this assumption.
Thus, in the remaining part of this proof, we concentrate on proving
 Eq.(\ref{eq sec sep x epsilon zenkashiki}) for $l=\eta$.

Here, we prove  Eq.(\ref{eq sec sep x epsilon zenkashiki}) 
for all $\eta \le l \le d$ by induction
starting from $l=d$. 
For $l=d$, Eq.(\ref{eq sec sep x epsilon zenkashiki}) trivially holds.
Suppose Eq.(\ref{eq sec sep x epsilon zenkashiki}) holds for $l$ satisfying
$1 \le \eta < l \le d$. 
 Then, from the definition of $\eta$, we have $\epsilon_l <
 \braket{\phi_l}{\psi_l}$, $\ket{\psi_l} \neq \ket{\phi_l}$, and
 $\braket{l}{\phi'_l}<0$.
Thus, from Lemma \ref{lemma sec sep zennkashiki},
there exists a state $\ket{\phi} \in {\rm span} \{ \ket{i} \}_{i=1}^l$
 satisfying $\braket{l}{\phi}=0$ and attaining the optimum of
 $X_{\epsilon_l}(\ket{\psi_l})$,
which is define by the optimization problem only on ${\rm span}\{ \ket{i} \}_{i=1}^l$.
Thus, in this case $X_{\epsilon_l}(\ket{\psi_l})$
can be rewritten as 
\begin{align}
\quad&  X_{\epsilon_l}(\ket{\psi_l})\nonumber \\
 =& \Big [ \max \big \{ 
\braket{\psi_l}{\phi} \ \big | \ \ket{\phi} \in {\rm span}\{ \ket{i} \}_{i=1}^{l-1}, \| \ket{\phi}\|^2\ge 1, \nonumber \\
\quad & \qquad  \forall i, \braket{i}{\phi}\ge \braket{i+1}{\phi} \ge
 0, \braket{\phi_l}{\phi}\le \epsilon_l   \big \} \Big ]^2
 \nonumber \\
=& \frac{\sum_{i=1}^{l-1}\lambda_i}{\sum_{i=1}^l\lambda_i} \cdot \Big [ \max \big \{ 
\braket{\psi_{l-1}}{\phi} \ \big | \ \ket{\phi} \in {\rm span}\{ \ket{i} \}_{i=1}^{l-1},  \nonumber \\
\quad & \qquad  \| \ket{\phi}\|^2\ge 1, \forall i, \braket{i}{\phi}\ge \braket{i+1}{\phi} \ge
 0,
 \nonumber \\
\quad & \qquad \braket{\phi_{l-1}}{\phi}\le \epsilon_{l-1}   \big \}
 \Big ]^2 \nonumber \\
=& \frac{\sum_{i=1}^{l-1}\lambda_i}{\sum_{i=1}^l\lambda_i} \cdot  X_{\epsilon_{l-1}}(\ket{\psi_{l-1}}),
\label{eq sec sep x epsilon zenkashiki 2}
 \end{align}
where we used relations
 $\sqrt{l}\braket{\phi_l}{\phi}=\sqrt{l-1}\braket{\phi_{l-1}}{\phi}$ and
$\sqrt{\sum_{i=1}^l\lambda_i}\braket{\psi_l}{\phi}=\sqrt{\sum_{i=1}^{l-1}\lambda_i}\braket{\psi_{l-1}}{\phi}$
 in the second equality.
Thus, from Eq.(\ref{eq sec sep x epsilon zenkashiki 2}) and Eq.(\ref{eq
 sec sep x epsilon zenkashiki}) for $l$, we derive Eq.(\ref{eq sec sep x
 epsilon zenkashiki}) for $l-1$.
Therefore, Eq.(\ref{eq sec sep x epsilon zenkashiki}) holds for all
 $\eta \le l \le d$.
\hfill $\square$
\end{Proof}

\section{Hypothesis testing under separable operations}\label{sec separable}
In this section, we treat the local hypothesis testing under separable
POVM, and gives a proof of Theorem \ref{main theorem sep}.
As we have predicted in the last section, the proof is completed by
showing that the local hypothesis testing under separable POVM is essentially
equivalent to the global
hypothesis testing treated in the last section, which is simpler than
the former. 

The equivalence of these two hypothesis testing problems can be
written as the following theorem in terms of their optimal success
probabilities $X_{\epsilon}(\ket{\psi})$ and 
$S_{\alpha, Sep}(\ket{\Psi})$:
\begin{Theorem} \label{lemma sec sep x epsilon = s delta sep}
For a state $\ket{\Psi}= \sum _{i=1}^{d_A}\sqrt{\lambda_i}\ket{ii} \in
 \Hi_{AB}$
and a state $\ket{\psi}=\sum _{i=1}^{d_A}\sqrt{\lambda_i}\ket{i} \in \Hi_A$,
\begin{equation}\label{eq sec sep x epsilon = s delta sep}
 S_{\alpha, Sep}(\ket{\Psi}) = X_{\sqrt{\alpha d_B}}(\ket{\psi}). 
\end{equation} 
\end{Theorem}
Since we have already derived an analytical formula for $X_{\sqrt{\alpha
d_B}}(\ket{\psi})$ in Theorem \ref{theorem main sec sep 1} of the last
section, 
we can derive Theorem \ref{main theorem sep} 
by just substituting $S_{\alpha, Sep}(\ket{\Psi})= 1-\beta_{\ket{\Psi}, Sep}(\alpha)$ 
instead of $X_{\sqrt{\alpha d_B}}(\ket{\psi})$ in 
Theorem \ref{theorem main sec sep 1}.
Therefore, a proof of Theorem \ref{main theorem sep} completely
reduces to a proof of Theorem \ref{lemma sec sep x epsilon = s delta
sep}.
Thus, we will concentrate on a proof of Theorem 
\ref{lemma sec sep x epsilon = s delta sep} in all the remaining part of
this section. The proof of this theorem can be divided into two parts:
In the first part, we show that $X_{\sqrt{\alpha d_B}}(\ket{\psi})$ is 
an upper bound of $S_{\alpha, Sep}(\ket{\Psi})$, and, then, in the
second part, we show that this upper bound is actually achievable by a separable POVM. 
Organization of this section is as follows: 
In the subsection A, we give an upper bound 
on $S_{\alpha, Sep}(\ket{\Psi})$, and show that
the separability condition of POVM in its definition can be replaced
by a condition in terms of a function $\chi(\rho)$.
Then, we investigate properties of $\chi(\rho)$ in subsection B.
Finally, in the subsection C, we complete the proof of Theorem \ref{lemma sec sep x epsilon =
s delta sep} by using lemmas derived in the subsection A and B.

\subsection{Reduction of the problem by means of a twirling}
In this subsection, we derive an upper bound of $S_{\alpha,
Sep}(\ket{\Psi})$ by using a twirling, which is a well-known technique to reduce a number of parameters
of an optimization problem in quantum information \cite{W89,BBPSSW96,BDSW96,R01,VW01}. 
Here, we use the twirling operation introduced in the paper \cite{OH08}.
Without losing generality, we can choose a computational basis as the
Schmidt basis of $\ket{\Psi}$ so that $\ket{\Psi}$ can be written as 
\begin{equation}
 \ket{\Psi}=\sum _{i=1}^{d_A}\sqrt{\lambda_i} \ket{ii},
\end{equation}
where $\{ \lambda _i \}_{i=1}^{d_A}$ is the Schmidt coefficients of $\ket{\Psi}$.
 We define a
family of local unitary operators $U_{\overrightarrow{\theta}}$ parametrized by
$\overrightarrow{\theta} = \{\theta _i \}_{i=1}^d$ as follows,
\begin{equation}\label{definition of U_theta}
U_{\overrightarrow{\theta}} = (\sum_{j=1}^{d_A} e^{i \theta _j} \ket{j}\bra{j})
\otimes (\sum _{k=1}^{d_A} e^{-i \theta _k} \ket{k}\bra{k} ).
\end{equation}
Note that $\left (\Hi_{AB},  U_{\overrightarrow{\theta}}  \right )$ 
is a unitary representation of the compact topological group $\overbrace{U(1) \times \cdots \times U(1)}^{d_A} $;
by means of a unitary representation of a compact topological group, 
we implement the "{\it twirling}" operation (the averaging  over the
compact topological group) for a state (or POVM) \cite{HMMOS07}. 
We write this twirling operation as $\Gamma$.   
Since by an action of twirling operation, a given state is projected to the subspace of all invariant elements of the group action \cite{HMMOS07},
 we can calculate $\Gamma(T)$ for any operator $T \in \B(\Hi_{AB})$ as follows:
\begin{align*}
\quad& \Gamma(T) \\
 \stackrel{\rm{ def} }{=} & \int  _0^{2 \pi} \cdots \int _0^{2
\pi} U_{\overrightarrow{\theta}} T U_{\overrightarrow{\theta}}^{\dagger} d \theta _1 \cdots d \theta _d \\
=& (\sum _{j=1}^d \ket{e_j}\bra{e_j} \otimes \ket{f_j}\bra{f_j} )T ( \sum _{j=1}^d \ket{e_j}\ket{e_j} \otimes \ket{f_j}\bra{f_j}) \\
\ & \quad +\sum _{j \neq k} \left ( \ket{e_j}\bra{e_j}\otimes
			      \ket{f_k}\bra{f_k} \right ) 
T \left ( \ket{e_j}\bra{e_j}\otimes \ket{f_k}\bra{f_k} \right ).
\end{align*}
Suppose $Q$ is a {\it maximally correlated subspace} with respect to the
computational basis: 
\begin{equation}
Q \stackrel{\rm def}{=} {\rm span} \{ \ket{ii} \}_{i=1}^{d_A}.
\end{equation} 
Then, the above equation guarantees that all states on $Q$
including $\ket{\Psi}$ are invariant
under the action of $\Gamma$:
\begin{equation}\label{inv rho on Bq}
\rho \in \B(Q) \Longrightarrow \Gamma(\rho) = \rho.
\end{equation}
Here, we note that every state $\rho $ on $Q$ is a so called  
{\it maximally correlated state} \cite{R01,HH04}.

Defining $\overline{S}_{\alpha, Sep}(\ket{\Psi}) $ as 
\begin{align}\label{eq sec sep def overline s delta sep}
 \quad & \overline{S}_{\alpha, Sep}(\ket{\Psi}) \nonumber \\ 
\stackrel{\rm def}{=}& \max \Big \{ \bra{\Psi}T\ket{\Psi} | \Tr T \le
			     \alpha d_A d_B, 0\le T \le I, T \in SEP \Big \},
\end{align}
where $SEP$ is the set of all (positive) separable operators on $\Hi$,
we can easily see 
\begin{equation}\label{eq sec sep s delta sep le overline s delta sep}
S_{\alpha, Sep}(\ket{\Psi}) \le  \overline{S}_{\alpha, Sep}(\ket{\Psi}). 
\end{equation}
We define a function $\chi(\rho)$ for a positive operator $\rho \in \mathcal{P}_{+} (Q)$ as 
\begin{align}\label{eq sec sep def chi rho}
 \chi(\rho)
\stackrel{\rm def}{=}& \min \{ \Tr \left ( \rho + \sigma \right )|
\exists \sigma = \sum _{j \neq k} p_{jk}\ket{j}\bra{j}\otimes
\ket{k}\bra{k}, \nonumber \\
\quad & \quad 0 \le \sigma \le I, \rho + \sigma \in SEP  \}. 
\end{align}
Then, we can show the following lemma:
\begin{Lemma}  \label{lemma sec sep overline S = max}
\begin{align}\label{eq lemma sec sep overline S = max}
\overline{S}_{\alpha, Sep}(\ket{\Psi}) =& \max \Big \{ \bra{\Psi}T_0
\ket{\Psi} \big |\ T_0 \in \B(Q),\nonumber \\
\quad& \quad \chi (T_0)\le\alpha d_A d_B,\  0 \le T_0 \le I_Q   \Big \},
\end{align}
where $I_Q$ is an identity operator of the space $Q$: $I_Q
 \stackrel{\rm def}{=} \sum _{i=1}^{d_A} \ket{ii}\bra{ii}$.
\end{Lemma}

\begin{Proof}
Suppose $T \in \B (\Hi_{AB})$ is optimal for $\overline{S}_{\alpha,
 Sep}(\ket{\Psi})$.
Then, from (\ref{inv rho on Bq}), we can easily show $\Gamma(T)$ is also optimal.
On the other hand, $\Gamma(T)$ can be written as 
\begin{equation}
 \Gamma(T) = T_0 + \sigma,
\end{equation}
where $T_0 \in \B (Q)$ and $\sigma$ can be written as 
$ \sigma = \sum _{j \neq k} p_{jk}\ket{j}\bra{j}\otimes
\ket{k}\bra{k}$.
Thus, we have 
\begin{align}
 \quad & \overline{S}_{\alpha, Sep}(\ket{\Psi}) \nonumber \\
=& \max \Big \{ \bra{\Psi}T_0
						\ket{\Psi}  \big |\  T_0 \in \B
						(Q),\  0 \le T_0 \le I_Q \nonumber \\
\quad & \quad  \sigma = \sum _{j \neq k} p_{jk}\ket{j}\bra{j}\otimes
\ket{k}\bra{k},\  0 \le \sigma \le I \nonumber \\
\quad & \quad T_0 + \sigma \in SEP,\  \Tr T_0+\sigma \le \alpha d_A d_B
 \Big \}\nonumber \\
 =&  \max \Big \{ \bra{\Psi}T_0
						\ket{\Psi} \big | \ T_0 \in \B
						(Q),\nonumber \\
\quad& \quad  \chi (T_0)\le\alpha d_A d_B,\  0 \le T_0 \le I_Q   \Big \}
\end{align}
\hfill $\square$
\end{Proof}

\subsection{Properties of a function $\chi (\rho)$}
In the previous subsection, we saw that 
$\overline{S}_{\alpha, Sep}(\ket{\Psi})$ gave an upper bound 
on $S_{\alpha, Sep}(\ket{\Psi})$,
and we can replace the separability condition of POVM in its definition
by a condition in terms of a function $\chi(\rho)$ defined as Eq.(\ref{eq sec sep def chi rho}).
For the purpose of further reduction of an upper bound, 
we give several important properties of $\chi(\rho)$ which we will
 use in the next subsection. 

First, $\chi(\rho)$ is closely related to an entanglement
 measure so called the robustness of entanglement \cite{VT99,HN03}. 
The robustness of entanglement is defined as
\begin{equation}
R_{s(g)}(\rho) \stackrel{\rm def}{=} \inf \left \{ \Tr \sigma : \sigma + \rho \in
		      SEP, \sigma \in C \right \}, 
\end{equation}
where $C$ is $SEP$ for $R_s(\rho )$ (the separable robustness of entanglement), and
$\mathcal{P}_+(\Hi_{AB})$ for $R_g(\rho)$ (the global
robustness of entanglement), respectively.
By the definition, they satisfy $R_g(\rho ) \le R_s(\rho )$.
It is also known that for a pure state $\ket{\Psi} = \sum _{i}
\sqrt{\lambda _i}\ket{ii}$, 
\begin{equation}\label{eq r_s r_g pure}
 R_s(\ket{\Psi}\bra{\Psi})= R_g(\ket{\Psi}\bra{\Psi})=\sum _{j \neq
  k}\sqrt{\lambda_j \lambda _k}.
\end{equation}
Generally, $\chi(\rho)$ gives an upper bound for $R_s(\rho)$ as follows:
\begin{Lemma}\label{lemma sep ine r_s chi}
 For all $\rho \in \mathcal{P}_+(Q)$,
\begin{equation}\label{eq r_s le chi - 1}
 R_s(\rho/\Tr \rho) \le \chi(\rho )/\Tr {\rho} -1. 
\end{equation}
\end{Lemma}
\begin{Proof}
By the definition, we have 
\begin{equation}\label{eq linearity chi}
  \chi \left (\rho/ \Tr \rho \right )=\chi (\rho)/ \Tr \rho.
\end{equation}
Suppose $\sigma$ attains the minimum of $\chi \left (\rho / \Tr \rho
 \right )$.
Then, since $\sigma$ is separable, 
\begin{align*}
R_s(\rho / \Tr \rho) \le & \chi \left (\rho / \Tr \rho \right ) - 1 \\
=& \chi(\rho)/ \Tr \rho
 -1 .
\end{align*}
\hfill $\square$
\end{Proof}
Moreover, for a pure state, we can prove the equality 
of Eq.(\ref{eq r_s le
chi - 1}); that is,   $\chi(\ket{\Psi}\bra{\Psi})$ is nothing but
equal to the robustness of entanglement $R_{s(g)}(\ket{\Psi}\bra{\Psi})$
except a constant term:
\begin{Lemma}\label{lemma analytic formula chi pure}
 For a non-normalized state $\ket{\Psi}=\sum_i a_i \ket{ii} \in Q$,
 \begin{equation}\label{eq analytic formula chi pure}
   \chi(\ket{\Psi}\bra{\Psi})=\sum_{jk}|a_j||a_k|.
 \end{equation}
Thus, for a normalized pure state $\ket{\Psi} \in Q$, 
\begin{equation}
  \chi(\ket{\Psi}\bra{\Psi})-1=R_s(\ket{\Psi}\bra{\Psi})=R_g(\ket{\Psi}\bra{\Psi})
\end{equation}
\end{Lemma}
\begin{Proof}
 First, we assume $\ket{\Psi} \in Q$ to be a pure state.
Since the Schmidt coefficients of $\ket{\Psi}$ are $\{ |a_i|
 \}_{i=1}^{d_A}$,
Lemma \ref{lemma sep ine r_s chi} and Eq.(\ref{eq r_s r_g pure})
 guarantee
\begin{equation}\label{eq r_s le chi}
 R_s(\ket{\Psi}\bra{\Psi})=\sum_{j \neq k } |a_j||a_k| \le \chi (\ket{\Psi}\bra{\Psi})-1.
\end{equation}
We define a new basis $\{ \ket{\tilde{i}}\}_{i=1}^{d_A}$ of $\Hi_A$ so that $\ket{\Psi}$ can be
 written down as $\ket{\Psi}=\sum _{i}|a_i| \ket{\tilde{i}i}$.
We also define $T_1$ and $\sigma$ as 
\begin{align}
 T_1 \stackrel{\rm def}{=}& \ket{a}\bra{a}\otimes \ket{b}\bra{b} \\
\sigma  \stackrel{\rm def}{=}& \sum_{j \neq k}
 |a_j||a_k|\ket{\tilde{j}}\bra{\tilde{j}}\otimes \ket{k}\bra{k},
\end{align}
where $\ket{a} \stackrel{\rm def}{=} \sqrt{|a_i|}\ket{\tilde{i}}$
and $\ket{b} \stackrel{\rm def}{=} \sqrt{|a_i|}\ket{i}$.
Straightforward calculations yield
\begin{equation}
 \ket{\Psi}\bra{\Psi}+\sigma =\Gamma(T_1) \in SEP.
\end{equation}
Thus, the definition of $\chi (\rho)$ implies
\begin{equation}
 \chi (\ket{\Psi}\bra{\Psi} ) \le \sum _{j \neq k} |a_j||a_k| + 1 = \sum _{jk}|a_j||a_k|.
\end{equation}
From the above inequalities and Eq.(\ref{eq r_s le chi}), 
an arbitrary normalized pure state $\ket{\Psi} \in Q$ satisfies
\begin{equation}
 \chi (\ket{\Psi}\bra{\Psi}) =\sum _{jk}|a_j||a_k|.
\end{equation}
Finally, by Eq.(\ref{eq linearity chi}), we can conclude the above equality hold
for all non-normalized pure states $\ket{\Psi} \in Q$, too.
\hfill $\square$
\end{Proof}

By using Lemma \ref{lemma sep ine r_s chi} and Lemma \ref{lemma analytic
formula chi pure},
we can prove that for a general mixed $\rho \in \mathcal{P}_+ (Q)$, 
$\chi (\rho)$ is derived by just a convex-roof extension from 
$\chi(\ket{\Psi}\bra{\Psi})$, which has analytic formula Eq.(\ref{eq
analytic formula chi pure}):
\begin{Lemma}\label{lemma sec sep eq chi = conv chi}
For $\rho \in \mathcal{P}_+ (Q)$,
 \begin{equation} \label{eq chi = conv chi}
  \chi(\rho) = \min _{\{p_i,\ket{\Psi_i} \}} \left \{ \sum _i p_i \chi
   (\ket{\Psi_i}\bra{\Psi_i})\Big  |\  \rho = \sum _i p_i
   \ket{\Psi_i}\bra{\Psi_i} \right \}.
 \end{equation}
\end{Lemma}
\begin{Proof}
 We first prove 
\begin{equation}\label{eq chi le conv chi}
 \chi (\rho) \le \min _{\{p_i,\ket{\Psi_i} \}} \left \{ \sum _i p_i \chi
   (\ket{\Psi_i}\bra{\Psi_i})\Big  |\  \rho = \sum _i p_i
   \ket{\Psi_i}\bra{\Psi_i} \right \}.
\end{equation}
Suppose $\rho$ can be decomposed as $\rho=\sum _i p_i
 \ket{\Psi_i}\bra{\Psi _i}$,
and $\sigma _i$ attains the minimum of $\chi (\ket{\Psi _i}\bra{\Psi _i})$;
that is, $\ket{\Psi_i}\bra{\Psi_i} + \sigma _i \in SEP$, $\chi
 (\ket{\Psi _i}\bra{\Psi _i})=\Tr (\ket{\Psi _i}\bra{\Psi _i} +\sigma
 _i)$,
and $\sigma _i$ also satisfies the remaining conditions.
Then, by defining $\sigma \stackrel{\rm def}{=} \sum _i p_i \sigma _i$, 
we have 
\begin{equation}
 \rho + \sigma = \sum _i p_i \left ( \ket{\Psi_i}\bra{\Psi_i}+ \sigma _i
			      \right ) \in SEP.
\end{equation}
It is also easy to check that $\sigma$ satisfies the remaining conditions
related to $\chi (\rho)$. 
Hence, we have $\chi (\rho) \le \sum _i p_i \chi (\ket{\Psi _i}\bra{\Psi
 _i})$. Therefore, the inequality (\ref{eq chi le conv chi}) holds.

Second, we prove
\begin{equation}\label{eq chi ge conv chi}
 \chi (\rho) \ge \min _{\{p_i,\ket{\Psi_i} \}} \left \{ \sum _i p_i \chi
   (\ket{\Psi_i}\bra{\Psi_i})\Big  |\  \rho = \sum _i p_i
   \ket{\Psi_i}\bra{\Psi_i} \right \}.
\end{equation}
Suppose $\sigma_{op}$ is optimal for $\chi(\rho)$. 
Then, since $\rho + \sigma _{op} \in SEP$, there exists an ensemble of
 pure states $\{ p_k, \ket{\xi _k} \}_k$ such that 
\begin{equation}\label{eq rho sigma p_k ket xi_k}
 \rho + \sigma _{op} = \sum _k p_k \ket{\xi _k} \bra{\xi _k}
\end{equation}
Since $\rho \in \mathcal{P}_+ (Q)$, $Q= span \{ \ket{ii} \}_{i=1}^{d_A}$, and
$\sigma _{op}$ can be written down as 
$\sigma _{op}=\sum _{i\neq j} q_{ij}\ket{i}\bra{i}\otimes
 \ket{j}\bra{j}$,
we can see that either $\ket{\xi_k} \in Q$ or there exist $i \neq j$
 satisfying $\ket{\xi _k } \propto
 \ket{ij}$. 
Thus, we can write $\rho + \sigma _{op}$ as
\begin{align}
 \quad & \rho + \sigma _{op} \nonumber \\
=& \sum _k p_k \Big  \{   \left ( \sum _i
					     \ket{ii}\bra{ii} \right )
\ket{\xi _k }\bra{\xi _k}\left ( \sum _i
					     \ket{ii}\bra{ii} \right )
					     \nonumber \\
\quad& \quad + \sum _{i \neq j} \left ( \ket{ij}\bra{ij} \right)\ket{\xi _k }\bra{\xi _k}\left ( \ket{ij}\bra{ij} \right)
\Big \}
\end{align}
Hence, defining  $\ket{\Psi _k}$ as
\begin{equation}
\ket{\Psi _k} = \left ( \sum _i
					     \ket{ii}\bra{ii} \right )
\ket{\xi _k },
\end{equation}
we derive $\rho=\sum _k p_k \ket{\Psi _k}\bra{\Psi _k}$.
Then, we can evaluate $\Tr \ket{\xi _k}\bra{\xi _k}$ as
\begin{align}
\quad & \Tr \ket{\xi _k}\bra{\xi _k} \nonumber\\
 =& \Tr \Gamma \left ( \ket{\xi
					       _k}\bra{\xi _k} \right
					       )\nonumber \\
=& \Tr \left ( \ket{\Psi_k}\bra{\Psi_k} +  \left ( \sum _{i \neq j}
					     \ket{ij}\braket{ij}{\xi_k}\braket{\xi_k}{ij}
					       \bra{ij} \right ) \right
					       )\nonumber \\
\ge & \chi (\ket{\Psi _k}\bra{\Psi _k}), \label{eq tr xi ge chi }
\end{align}
where we used the fact $\Gamma ( \ket{\xi _k}\bra{\xi _k}) \in SEP$ in
 the third line.
Thus, we can evaluate $\chi (\rho)$ as follows:
\begin{align}
 \chi (\rho ) =& \Tr ( \rho + \sigma _{op}) \nonumber \\
=& \sum _k p_k \Tr \ket{\xi _k}\bra{\xi _k} \nonumber \\
\ge & \sum _k p_k \chi (\ket{\Psi _k}\bra{\Psi_k}),
\end{align}
where we used Eq.(\ref{eq rho sigma p_k ket xi_k}) in the second line
 and the inequality (\ref{eq tr xi ge chi }) in the third line.
The above inequality guarantees that 
the inequality (\ref{eq chi ge conv chi}) holds.
From the inequalities (\ref{eq chi ge conv chi}) and (\ref{eq chi le conv chi}), 
Eq.(\ref{eq chi = conv chi}) holds.
\hfill $\square$
\end{Proof}

As the next step, for an operator $\rho \in \mathcal{P}_+(Q)$, we define a new function $\chi'(\rho)$ as
\begin{equation}
 \chi'(\rho ) = \sum _{ij}|\beta _{ij}|,
\end{equation}
where the coefficients $\{ \beta _{ij} \}_{ij}$ are defined as 
$\rho = \sum _{ij}\beta_{ij}\ket{ii}\bra{jj}$.
Then, we can show $\chi'(\rho)$ is a lower bound of $\chi(\rho)$:
\begin{Lemma} \label{lemma sec sep chi ge chi'}
For $\rho \in \mathcal{P}_+(Q)$,
 \begin{equation}
  \chi(\rho) \ge \chi'(\rho).
 \end{equation} 
Moreover, if ${\rm rank} \rho=1$, the equality holds.
\end{Lemma}
\begin{Proof}
 Suppose $\rho = \sum _k p_k \ket{\Psi_k}\bra{\Psi_k}$ is a decomposition which attains $\chi(\rho)$,
and $\{ a_i^{(k)}\}_{ik}$ are  coefficients defined as $
 a_i^{(k)} \stackrel{\rm def}{=}\braket{ii}{\Psi _k}$.
Then, we can evaluate $\chi(\rho)$ as follows:
\begin{align*}
 \chi (\rho) = & \sum _k p_k \chi (\ket{\Psi_k}\bra{\Psi_k}) \nonumber \\
=& \sum _k p_k \sum _{ij}|a_i^{(k)}a_j^{(k)}| \nonumber \\
=& \sum _k p_k \sum _{ij} |\braket{ii}{\Psi_k}\braket{\Psi_k}{jj}|
 \nonumber \\
 \ge & \sum _{ij} | \sum _k p_k \braket{ii}{\Psi_k}\braket{\Psi_k}{jj}|
 \nonumber \\
=& \sum _{ij} |\bra{ii}\rho \ket{jj}|\nonumber \\
=& \chi'(\rho),
 \end{align*}
where we used Eq.( \ref{eq chi = conv chi}) in the first line.
Moreover, when ${\rm rank}\rho=1$, we can easily see $\chi(\rho)=\chi'(\rho)$ from
 Lemma \ref{lemma analytic formula chi pure}.
\hfill $\square$
\end{Proof}

\subsection{Proof of Theorem }
In this subsection,
by using lemmas derived in the previous subsections,
we complete a proof of Theorem 
\ref{lemma sec sep x epsilon = s delta sep}.

First, by defining a new function
$\overline{S}'_{\alpha}(\ket{\Psi})$
as 
\begin{align}
\quad& \overline{S}'_{\alpha}(\ket{\Psi})\nonumber \\ 
= & \max_{T} \{ \bra{\Psi}T\ket{\Psi} | T \in \B
  (Q), 0 \le T \le I_Q, \chi'(\rho) \le \alpha d_A d _B \},
\end{align}
The following inequality follows from Lemma \ref{lemma
sec sep overline S = max}
and Lemma \ref{lemma sec sep chi ge chi'}:
\begin{equation}\label{eq sec sep overline s' delta ge overline s delta sep}
 \overline{S}'_{\alpha}(\ket{\Psi}) \ge \overline{S}_{\alpha, Sep}(\Psi).
\end{equation}
Now, in the definition of $\overline{S}'_{\alpha}(\Psi)$, 
all related operators are spanned by $\{ \ket{ii} \}_{i=1}^{d_A}$,
and a condition related to separability no more appears. 
Therefore, we have the following lemma
\begin{Lemma}\label{lemma sec sep overline s' = max psi'}
 \begin{align}\label{eq sep overline s' = max psi'}
 \overline{S}'_{\alpha}(\ket{\Psi}) = & \max_T \{ \bra{\psi}T\ket{\psi} | T \in \B
  (\Hi_A),  0 \le T \le I_A,\nonumber \\
\quad & \quad \quad \sum_{ij}|\bra{i}T\ket{j}| \le \alpha d_A d _B \},
\end{align}
where $\ket{\psi}$ is defined as $\ket{\psi} \stackrel{\rm def}{=}
 \sum _i \sqrt{\lambda _i}\ket{i}$ with the Schmidt coefficients $\{
 \lambda _i \}_i$ of $\ket{\Psi}$.
\end{Lemma}
Moreover, we can restrict a POVM element $T$ to a real operator.
\begin{Lemma}\label{lemma sec sep overline S' = max Psi' T=ReT}
  \begin{align}\label{eq overline S' = max Psi' T=ReT}
 \overline{S}'_{\alpha}(\ket{\Psi}) = & \max_T \{ \bra{\psi}T\ket{\psi} | T \in \B
  (\Hi_A),  0 \le T \le I_A,\nonumber \\
\quad & \quad \quad T= {\rm Re}T, \  \sum_{ij}|\bra{i}T\ket{j}| \le \alpha d_A d _B \},
\end{align}
where ${\rm Re} T$ is defined as ${\rm Re} T \stackrel{\rm def}{=} \sum
 _{ij} {\rm Re} \bra{i}T \ket{j} \ket{i}\bra{j}$.
\end{Lemma} 
\begin{Proof}
 Suppose $T$ is an optimal operator attaining $\chi'(\rho)$ in Eq.(\ref{eq sep overline s' = max
 psi'}).
Then, $\overline{T}$ defined as $\overline{T}  \stackrel{\rm def}{=} \sum
 _{ij} \overline{ \bra{i}T \ket{j}}  \ket{i}\bra{j}$
is also optimal and attains $\chi'(\rho)$.
Thus,  defining $T' \stackrel{\rm def}{=} T + \overline{T}/2$, 
we derive $T'= {\rm Re} T'$, $0 \le T' \le I_A$, $\bra{\psi}T'\ket{\psi}$, and 
\begin{align}
  \sum _{ij}| \bra{i}T' \ket{j} | 
\le& \frac{1}{2} \left \{  \sum _{ij}|\bra{i}T\ket{j}|+ \sum
		   _{ij}|\bra{i}\overline{T}\ket{j}|\right
		  \}\nonumber \\
\le & \alpha d_A d_B.
\end{align}
Thus, we derive Eq.(\ref{eq overline S' = max Psi' T=ReT}).
\hfill $\square$
\end{Proof}

Now, we can show that $X_{\sqrt{\alpha d_B}}(\ket{\psi})$
is  an upper bound of $\overline{S}'_{\alpha}(\Psi)$:
\begin{Lemma} \label{lemma sec sep x epsilon ge s delta sep}
For a state $\ket{\Psi}= \sum _{i=1}^{d_A}\sqrt{\lambda_i}\ket{ii} \in
 \Hi_{AB}$
and a state $\ket{\psi}=\sum _{i=1}^{d_A}\sqrt{\lambda_i}\ket{i} \in \Hi_A$,
\begin{equation}\label{eq overline s' le overline s''}
 \overline{S}'_{\alpha}(\ket{\Psi}) \le X_{\sqrt{\alpha d_B}}(\ket{\psi}). 
\end{equation} 
\end{Lemma}
\begin{Proof}
  Observing that for $x_i \in \mathbb{R}$, 
\begin{equation}
 \sum _i |x_i| \le \epsilon \Longleftrightarrow 
\forall \vec{k} \in \mathbb{Z}_2^d, \sum _i (-1)^{k_i}x_i
  \le \epsilon,  
 \end{equation}
we can evaluate Eq.(\ref{eq overline S' = max Psi' T=ReT}) as
\begin{align}\label{eq overline S' = max Psi' T=ReT vec k}
\quad& \overline{S}'_{\alpha}(\ket{\Psi}) \nonumber \\ 
 = & \max_T \{ \bra{\psi}T\ket{\psi} | T \in \B
  (\Hi_A),  0 \le T \le I_A, T= {\rm Re}T, \nonumber \\
\quad & \quad \quad \forall \vec{k} \in \mathbb{Z}_2^{d_A \times d_A},
 \sum_{ij} (-1)^{k_{ij}}\bra{i}T\ket{j} \le \alpha d_A d _B
 \} \nonumber \\
\le& \max_T \{ \bra{\psi}T\ket{\psi} | T \in \B
  (\Hi_A),  0 \le T \le I_A,  T= {\rm Re}T, \nonumber \\
 \quad & \quad \quad \forall \vec{k}\in
 \mathbb{Z}_2^{d_A}, |\bra{\phi _{\vec{k}}}T\ket{\phi_{\vec{k}}}| \le \alpha d _B \},
\end{align}
where we use the observation $\bra{\phi_{\vec{k}}}T\ket{\phi _{\vec{k}}}= \frac{1}{d_A}
 \sum _{ij} (-1)^{k_i+k_j}\bra{i}T\ket{j}$ in the above inequality.
Then, by using the same argument of the proof of Lemma 
\ref{lemma sec sep overline S' = max Psi' T=ReT}, 
we can remove the restriction of positivity of $T$ from the maximization
in the  last line Eq.(\ref{eq overline S' = max Psi' T=ReT vec k}), and
 derive the inequality (\ref{eq overline s' le overline s''}).

\hfill $\square$
\end{Proof}

Finally, the inequalities (\ref{eq sec sep s delta sep le overline s delta sep}),
(\ref{eq sec sep overline s' delta ge overline s delta sep}),
(\ref{eq overline s' le overline s''}) yield
\begin{eqnarray}
 S_{\alpha, Sep}(\ket{\Psi})\le X_{\sqrt{\alpha d_B}}(\ket{\psi}).
\end{eqnarray} 
Thus, we have succeeded to prove that the optimal success probability of
the global hypothesis testing in the last section 
$X_{\sqrt{\alpha d_B}}(\ket{\psi})$ is an upper bound of the optimal
success probability of the local hypothesis testing $S_{\alpha, Sep}(\ket{\Psi})$.
Thus, in order to complete the proof of Theorem 
\ref{lemma sec sep x epsilon = s delta sep},
the remaining task is to show  the above inequality is actually an equality.
This can be done as follows:
\begin{Proof}[Theorem \ref{lemma sec sep x epsilon = s delta sep}] 
The inequalities (\ref{eq sec sep s delta sep le overline s delta sep}),
(\ref{eq sec sep overline s' delta ge overline s delta sep}) and
(\ref{eq overline s' le overline s''}) yield
\begin{align}
 S_{\alpha, Sep}(\ket{\Psi}) \le &
\overline{S}_{\alpha,  Sep}(\ket{\Psi}) \nonumber \\
\le&
\overline{S}_{\alpha}'(\ket{\Psi}) \nonumber \\
\le& X_{\sqrt{\alpha d_B}}(\ket{\psi}).
\end{align} 
Thus, we just need to show that the above three inequalities are
 actually equalities.

First, we prove the equality
\begin{equation}\label{eq sec sep overline s delta sep' = overline s
 delta sep''}
 \overline{S}_{\alpha}'(\ket{\Psi}) 
= X_{\sqrt{\alpha d_B}}(\ket{\psi}).
\end{equation}
From Theorem \ref{theorem main sec sep 1}, 
an optimal operator $T \in \Hi _A$ of $X_{\sqrt{\alpha d_B}}(\ket{\psi})$
can satisfy the condition $\bra{i}T\ket{j}\ge0$ for all $i$ and $j$.
Hence, we can calculate as
\begin{align}
 \sum _{ij}|\bra{i}T\ket{j}| =&  \sum _{ij}\bra{i}T\ket{j} \nonumber\\
=& d_A \bra{\phi_d}T\ket{\phi_d}\nonumber \\
\le & \alpha d_A d_B.
\end{align}
Thus, from Lemma \ref{lemma sec sep overline S' = max Psi' T=ReT}, 
we derive the equality (\ref{eq sec sep overline s delta sep' = overline s
 delta sep''}).

Second, we prove the equality
\begin{equation}\label{eq sec sep overline s delta sep = overline s
 delta sep'}
 \overline{S}_{\alpha, Sep}(\ket{\Psi}) 
= \overline{S}_{\alpha}'(\ket{\Psi}).
\end{equation}
Suppose $T \in \B (Q)$ attains the optimum of
 $\overline{S}_{\alpha}'(\ket{\Psi})$; thus, 
$T$ satisfies $0 \le T \le I_Q$ and $\chi'(T) \le \alpha d_A d_B$.
From Theorem \ref{theorem main sec sep 1}, 
an optimal $T$ can be written as $T= \ket{\Phi}\bra{\Phi}$; that is, 
${\rm rank}T=1$.
Thus,  from Lemma \ref{lemma sec sep chi ge chi'},
we derive $\chi(T)=\chi'(T)$.
This fact and Lemma  \ref{lemma sec sep overline S = max} guarantee the equality (\ref{eq sec sep overline s delta sep = overline s
 delta sep'}).

Finally, we prove the equality
\begin{equation}\label{eq sec sep s delta sep = overline s
 delta sep'}
 S_{\alpha, Sep}(\ket{\Psi}) 
= \overline{S}_{\alpha, Sep}(\ket{\Psi}).
\end{equation} 
First, an optimal operator $T$ attaining the optimum of
Eq.(\ref{eq sec sep def overline s delta sep}) can be written down
as 
\begin{equation}
 T=T_0+\sigma,
\end{equation}
where $T_0$ is an operator attaining the optimum of Eq.(\ref{eq lemma sec sep overline S = max})
and $\sigma$ is an operator attaining $\chi(T)$ in terms of 
Eq.(\ref{eq sec sep def chi rho}).

When $\eta=1$ in Theorem \ref{theorem main sec sep 1} for
 $d=d_A$ and $\epsilon = \sqrt{\alpha d_B}$,
we can choose $T_0$ as $T_0=\epsilon_2^2 \ket{11}\bra{11}$,
where $\epsilon_2 \stackrel{\rm def}{=}\min \{1, \sqrt{\alpha d_A d_B/2}\}$.
In this case, since $T_0$ is already separable, $\chi(T_0)=1$
and $\sigma=0$. Hence, $T=T_0=\epsilon_2^2 \ket{11}\bra{11}$. 
Thus, $I-T$ is a separable operator. Therefore, 
Eq.(\ref{eq sec sep s delta sep = overline s delta sep'}) holds.

Suppose $\eta \ge 2$ in Theorem \ref{theorem main sec sep 1} for
 $d=d_A$ and $\epsilon = \sqrt{\alpha d_B}$.
Then, $T_0$ can be chosen as a normalized pure state.
Thus, we have 
\begin{equation}
 \chi(T_0)=1+R_g(T_0).
\end{equation}  
Actually, it is known (in the proof of Lemma 1 of \cite{OH08}) that we can choose an optimal $\sigma$
as 
\begin{equation}
 \sigma = \sum _{j \neq k} 
\sqrt{\lambda _j \lambda _k}\ket{j}\bra{j}\otimes \ket{k}\bra{k},
\end{equation}
where $\{ \lambda_i \}_{i=1}^{d_A}$ is the Schmidt coefficient of the pure
 state $T_0$.
It has already proven that, if $T$ is defined as $T=T_0 + \sigma$ by
 using the above $\sigma$,  $I-T$ is also separable ( in the proof of
 Theorem 2 of \cite{OH08}).
Thus, 
Eq.(\ref{eq sec sep s delta sep = overline s delta sep'}) holds also in
 this case.
Therefore, Eq.(\ref{eq sec sep x epsilon = s delta sep}) holds.
\hfill $\square$
\end{Proof}

\section{Summary}\label{sec summary}
 In this paper, we have treated a local hypothesis testing whose alternative
hypothesis is a bipartite pure state $\ket{\Psi}$, and whose null
hypothesis is the completely mixed state.  
As a result, we have analytically derived an optimal type 2 error and 
an optimal POVM for one-way LOCC POVM (Theorem \ref{main theorem one-way})
and Separable POVM (Theorem \ref{main theorem sep}). In particular, in
order to derive an analytical solution for Separable POVM, we have  
proved the equivalence of the local hypothesis testing under
Separable POVM and a global hypothesis testing with a composite
alternative hypothesis (Section \ref{sec separable}), 
and analytically solved this global hypothesis 
testing (Section \ref{sec global}).
Furthermore, for two-way LOCC POVM, we have studied a family of
simple three-step LOCC protocols, and have showed that the best 
protocol in this family has strictly better performance than any one-way
LOCC protocol in all the cases where there may exist
difference between  two-way LOCC POVM and one-way LOCC POVM (Section
\ref{sec locc}).  

 Although we restrict ourselves on treating the hypothesis-testing problem in a single-copy
 scenario in this paper, we are also interested in an extension of our
 results to problem settings with asymptotically infinite
 copies of the hypotheses, that is, problem settings like
 Stein's Lemma \cite{HP91}, and the Chernoff bound \cite{ACMBMAV07}.  
In particular, it is interesting whether the difference of optimal error
 probabilities under one-way  and two-way LOCC survives in the
 asymptotic extension of the problem. 
Actually, we have derived new results on this asymptotic extension and are on the way to prepare a manuscript \cite{OH10}.

\section*{Acknowledgment}
The authors would like to tank Prof. K. Nemoto for discussion which
motivates this study. 
MO are supported by JSPS Postdoctoral Fellowships for Research Abroad.
MH was also partially supported by a MEXT Grant-in-Aid for
Young Scientists (A) No. 20686026. The Center for Quantum Technologies
is funded by  the Singapore Ministry of Education and the National
Research Foundation  as part of the Research Centres of Excellence programme.

\appendices
\section{Proof of Statements}
\subsection{Proof of Corollary \ref{corollary sec main}}\label{appendix
proof sec main corollary} 
The statement about a product state and a maximally entangled state is
trivial from Theorem \ref{main theorem one-way}, \ref{main theorem two-way} and
\ref{main theorem sep}. Thus, we only give a proof about non-maximally
entangled states here.
Suppose $\alpha <1/d_Ad_B$, that is, $\epsilon _1 < 1$.
Then, for $l \ge 2$,
\begin{align}
 \braket{\phi_l}{\psi_l}-\epsilon_l =& \left (\frac{\sum _{i=1}^l
  \sqrt{\lambda _i}}{\sqrt{\sum_{i=1}^l\lambda_i}} - \epsilon_1 \right
  )/\sqrt{l} \nonumber \\
>& \left (\frac{\sum _{i=1}^l
  \sqrt{\lambda _i}}{\sqrt{\sum_{i=1}^l\lambda_i}} - 1 \right
  )/\sqrt{l} \nonumber \\
>&0.
\end{align}
Thus, $\braket{\phi_l}{\psi_l}>\epsilon_l$ for $l \ge 2$.

In the remaining part, we will prove the statement for separable POVM in the case $\lambda_1>\lambda_2$
and in the case $\lambda_1=\lambda_2$, separately.
\begin{enumerate}
 \item In the case $\lambda_1>\lambda_2$, we can prove $\braket{2}{\phi_2'}<0$ as
       follows:
\begin{align}
\quad&
 2\sqrt{2}\sqrt{(\lambda_1+\lambda_2)(1-c_2^2)}\braket{2}{\phi_2'}
 \nonumber \\
=& (\epsilon_1-\sqrt{2-\epsilon_1^2})(\lambda_1-\lambda_2)\nonumber \\
<&0.
\end{align}
Thus, $\epsilon_2<\braket{\phi_2}{\psi_2}$,
       $\ket{\psi_2}\neq\ket{\phi_2}$, and $\braket{2}{\phi_2'}<0$
       guarantee $\eta=1$. Thus, from Theorem \ref{main theorem sep},
we derive $\beta_{\alpha,sep}=1-\lambda_1\alpha d_Ad_B$ and the optimal POVM
       is given by $T(\epsilon_1\ket{0})=\alpha d_Ad_B\ket{00}\bra{00}.$
\item In the case $\lambda_1=\lambda_2$, there exists a number $\eta_0$ such that
       $\lambda_1=\cdots =\lambda_{\eta_0}>\lambda_{\eta_o+1}$. In this case, we can
       easily see $\eta=\eta_0$. Then, $c_{\eta_0}=1$ guarantees
\begin{align}
 \beta_{\alpha, sep}(\ket{\Psi})=& 1-\left
				       (\sum_{i=1}^{\eta_0}\lambda_i \right )\cdot
				       \frac{\alpha d_Ad_B}{\eta_0}
				       \nonumber \\
=& 1-\lambda_1\alpha d_Ad_B.
\end{align}
We can easily check that a POVM $T=\alpha d_Ad_B\ket{11}\bra{11}$
       attains this optimum.
\end{enumerate}
Finally, since the above POVM can be implemented by one-way LOCC, we
derive the statement of the corollary.
\hfill $\square$

\ifCLASSOPTIONcaptionsoff
  \newpage
\fi




\end{document}